\documentclass[aps,pre,floats,superscriptaddress,nofootinbib,floatfix,twocolumn,longbibliography,showkeys]{revtex4-1}
\usepackage{amssymb,amsmath,amsfonts,amsthm}
\usepackage{graphicx}
\usepackage{newtxtext}
\usepackage{dcolumn}
\usepackage{bm}
\usepackage{comment}
\usepackage{csquotes}
\usepackage{mathtools}
\usepackage{gensymb}
\usepackage{hyperref}
\usepackage{makecell}
\usepackage{changepage}
\usepackage{multirow}
\usepackage[table]{xcolor}
\usepackage[mathscr]{euscript}
\usepackage{ulem}
\graphicspath{{./fig/}}

\makeatletter
\AtBeginDocument{%
  \def\bibsection{%
    \par
    \onecolumngrid@push
    \vspace*{1cm}
    \begingroup
      \baselineskip26\p@
      \bib@device{\textwidth}{245.5\p@}%
    \endgroup
    \nobreak\@nobreaktrue
    \addvspace{19\p@}%
    \par
    \onecolumngrid@pop
  }%
}
\makeatother

\hypersetup{
	colorlinks=true,
	linkcolor=blue,
	filecolor=blue,      
	urlcolor=blue,
	citecolor=blue,
}

\begin{document}

\title{Field‐controlled interfacial transport and pinning in an active spin system}
\author{Mintu Karmakar$^{\star}$}
\email{mkarmakar094@ucas.ac.cn}
\affiliation{Wenzhou Institute of the University of Chinese Academy of Sciences, Wenzhou, Zhejiang 325011, China.}
\affiliation{School of Physical Sciences, University of Chinese Academy of Sciences, Beijing 100049, China.}
\affiliation{Departament de F\'isica de la Mat\`eria Condensada, Universitat de Barcelona, Mart\'i i Franqu\`es 1, E08028 Barcelona, Spain.}

\author{Matthieu Mangeat$^{\star}$}
\email{mangeat@lusi.uni-sb.de}
\affiliation{Center for Biophysics \& Department for Theoretical Physics, Saarland University, 66123 Saarbr{\"u}cken, Germany.}

\author{Swarnajit Chatterjee}
\email{swarnajit.chatterjee@cyu.fr}
\affiliation{Laboratoire de Physique Th{\'e}orique et Mod{\'e}lisation, UMR 8089, CY Cergy Paris Universit{\'e}, 95302 Cergy-Pontoise, France.}

\author{Heiko Rieger}
\email{heiko.rieger@uni-saarland.de}
\affiliation{Center for Biophysics \& Department for Theoretical Physics, Saarland University, 66123 Saarbr{\"u}cken, Germany.}

\author{Raja Paul}
\email{raja.paul@iacs.res.in}
\affiliation{School of Mathematical \& Computational Sciences, Indian Association for the Cultivation of Science, Kolkata -- 700032, India.}

\begin{abstract}
Field control provides a practical route to programmable active matter, yet how weak fields modify non-equilibrium coexistence and interfaces remains unclear. To address this, we study a minimal flocking model of active Potts particles coupled to an external field and show that even weak fields can reconfigure phase behavior and interfacial dynamics. For a homogeneous unidirectional field, the flocking phase is reshaped: the coexistence regime between an apolar gas and a polar liquid is replaced by a phase separation between two field-aligned polar phases: a low-density, weakly polarized background and a high-density, strongly polarized band, both moving along the field. When the system forms a dense longitudinal lane oriented transverse to the field, it executes a slow treadmilling motion against the field, driven by the weakly polarized background. If the system is divided into regions with opposite field directions, particles accumulate at the interface, leading to field-induced interface pinning with flocks performing back-and-forth oscillatory motion. In the presence of quenched random field orientations, this pinning favors a disordered state in which global order diminishes with increasing system size, consistent with Imry--Ma arguments, while the quenched disorder smoothens sharp first-order signatures, in line with the Aizenman--Wehr theorem, with activity modifying the scaling. A coarse-grained hydrodynamic theory supports these observations and is consistent with microscopic simulations.
\end{abstract}

\maketitle
\def\thefootnote{{$\star$}}\footnotetext{These authors have contributed equally to this work}\def\thefootnote{\arabic{footnote}}

\section{Introduction}
Systems of self-propelled particles, collectively known as active matter, represent a vibrant frontier of non-equilibrium statistical physics that has been studied extensively over the past three decades~\cite{marchetti2013hydrodynamics,de2015introduction,bechinger2016active,needleman2017active,shaebani2020computational,chate2020dry,bowick2022symmetry,hallatschek2023proliferating,gompper2025}. Spanning from bacterial colonies and cellular tissues to bird flocks and synthetic microswimmers~\cite{becco2006experimental,ballerini2008interaction,steager2008dynamics,cavagna2014bird,calovi2014swarming,gomez2022intermittent,deseigne2010collective,xi2019material,chen2024emergent}, these systems comprise agents that consume energy to generate directed motion~\cite{vicsek2012collective}. This continuous energy consumption at the microscopic level gives rise to flocking~\cite{toner2024physics}, a large-scale collective motion without an equilibrium counterpart, emerging from simple local alignment rules~\cite{vicsek1995novel,gregoire2004onset,chate2008collective,chate2008modeling}.

While much of the foundational work has focused on characterizing these spontaneous behaviors in idealized, homogeneous settings~\cite{vicsek1995novel,toner1995long,toner1998,gregoire2004onset}, real active systems---from cells navigating complex extracellular matrices and bacteria swimming in porous media to animal herds migrating through forests and complex landscapes---must contend with environmental heterogeneity~\cite{petrie2016multiple,bhattacharjee2019bacterial,ariel2022variability}. In statistical physics, this complexity is often modeled as quenched random disorder---spatially fixed randomness (obstacles, pinning fields) that remains static on the relevant timescale---which can profoundly impact collective order~\cite{chepizhko2013diffusion,reichhardt2014active,morin2017distortion,das2018ordering,toner2018swarming,kumar2020active,peruani2018cold,aranson2022confinement,rahmani2021topological,chen2022incompressible,jena2026competing}. Recently, beyond characterization, attention has shifted to control. External fields are being used to guide, sculpt, and program the self-organization of natural and synthetic active matter systems into desired functional states~\cite{norton2020optimal, davis2024control, zhang2022guiding, pellicciotta2023colloidal, yang2025dynamic, boymelgreen2022synthetic,rey2023light,palacios2021guided,activesolid2025}.

External fields also provide a complementary, versatile toolkit for manipulating active systems. Magnetic, electric, and optical fields have proven effective in directing transport, inducing pattern-forming instabilities, and assembling dynamic, tunable structures~\cite{koessel2019controlling, koessel2020emergent,zhang2022guiding,pellicciotta2023colloidal}. Examples include electrically powered Quincke rollers that realize field-programmable collective motion~\cite{bricard2013emergence}, light-enabled, feedback-controlled density and flow programming in photokinetic bacteria~\cite{frangipane2018dynamic}, field-modulated structures in active dipolar suspensions~\cite{parage2025modulation}, and optically defined boundaries that program active biomolecular gels~\cite{ross2019controlling}. In polar active fluids, imposed flows likewise act as aligning fields, producing hysteresis and geometry-protected states~\cite{morin2018response}. Beyond physical drives, agent-level information cues---ranging from leadership to broadcast communication---can also introduce a field-like bias that shapes collective behavior~\cite{pearce2016linear,ziepke2025acoustic}. From a theoretical standpoint, a uniform biasing field offers a clean theoretical probe that isolates the influence of a persistent, global symmetry-breaking cue from the intricacies of a disordered landscape~\cite{brambati2022signatures, ginelli2016leading}.

With a uniform biasing field as a simple, tunable control knob for orientational symmetry breaking~\cite{brambati2022signatures, ginelli2016leading}, we seek a minimal framework in which its coupling to collective orientation is explicit. There are two canonical routes to flocking in minimal models: (a) alignment of continuous orientations in the discrete- or continuous-time Vicsek framework~\cite{vicsek1995novel,solon2015phase,solon2015pattern,martin2018collective,fruchart2021non,chatterjee2023flocking}; and (b) discrete-symmetry framework of active spin models---active Ising (AIM)~\cite{solon2013revisiting,solon2015flocking,solon2015pattern,AIM2023,AIM2024,woo2024motility,mangeat2025emergent}, active Potts (APM)~\cite{chatterjee2020flocking, mangeat2020flocking,karmakar2023jamming,chatterjee2025stability}, and active clock (ACM)~\cite{chatterjee2022polar,solon2022susceptibility}. In particular, in this work, we focus on the 4-state APM~\cite{chatterjee2020flocking}, whose robust flocking phenomenology---including band–to–lane reorientation transitions~\cite{chatterjee2020flocking,mangeat2020flocking}, flocking–to–jamming transitions under volume exclusion~\cite{karmakar2023jamming}, and a non-reversal {\it sandwich} configuration arising from localized perturbations~\cite{chatterjee2025stability}---provides an ideal setting for a controlled study of field-driven reorientation, steering, and programming of a macroscopic active band.

\section{Model and Simulation details}

We consider an ensemble of $N$ particles defined on a two-dimensional square lattice of size $L^2$ with periodic boundary conditions applied on both sides, where $L$ is the linear lattice dimension. The average particle density $\rho_0$ is then defined as $\rho_0=N/L^2$. The model is built upon the $q$-state APM~\cite{chatterjee2020flocking, mangeat2020flocking} in which the dynamics are governed either by the on-site flipping of the internal spin state or by nearest-neighbor hopping.

Besides, we now applied an external magnetic field ${\bf h}_i$, placed at each site $i$ of the lattice. Each particle has a spin state $\sigma$ which takes an integer value in $[1,q]$. The Hamiltonian for the APM with $L^2$ sites can be expressed as~\cite{paul2002low} $H_{\rm APM} = \sum_{i} {\cal H}_i$ with
\begin{equation}
\label{Hapm}
{\cal H}_i=-\frac{J}{2\rho_i}\sum_{j=1}^{\rho_i}\sum_{k\ne j}\left(q\delta_{\sigma_i^j,\sigma_i^k}-1\right)- h_i\sum_{j=1}^{\rho_i}\left(q\delta_{\sigma_i^j,\alpha_i}-1\right) \, ,
\end{equation}
where $\delta_{\sigma,\sigma^\prime}$ is the Kronecker-delta ($\delta_{\sigma,\sigma^\prime}=1$ for $\sigma=\sigma^\prime$ and $\delta_{\sigma,\sigma^\prime}=0$ otherwise) and $\sigma_i^j$ is the spin state of the $j^{\rm th}$ particle on site $i$. $J$ is the strength of the ferromagnetic coupling between the particles, $h_i$ and $\alpha_i$ denotes the strength and the direction of the field at site $i$, respectively. $\rho_i$ is the number of particles at site $i$, defined by $\rho_i=\sum_{\sigma=1}^{q}n_{\sigma,i}$ where $n_{\sigma,i}$ is the number of particles in state $\sigma$ at site $i$, whereas, the local magnetization corresponding to state $\sigma$ at site $i$ is denoted by:
\begin{equation}
\label{Hapm_mag}
m_{\sigma,i}=\sum_{j=1}^{\rho_i}\frac{q\delta_{\sigma,\sigma_i^j}-1}{q-1}=\frac{qn_{\sigma,i}-\rho_i}{q-1} \, .
\end{equation}
The global magnetization in state $\sigma$ is denoted by $M_\sigma = \sum_i m_{\sigma,i} / N$. To quantify the overall breaking of the four-state rotational symmetry, we also define the scalar global order parameter:
\begin{equation}
\label{eqmag}
m = \sqrt{\frac{q-1}{q}\sum_{\sigma} M_\sigma^2} = \sqrt{\frac{q\sum_{\sigma} p_\sigma^2-1}{q-1}} \, .
\end{equation}
where $p_\sigma = N_\sigma/N$ and $N_\sigma$ is the number of particles in state $\sigma$ ($N=\sum_\sigma N_\sigma$). With this normalization, $m=1$ when all particles occupy a single state and $m=0$ when the four states are equally populated ($p_\sigma=1/q$).

A particle at site $i$ with state $\sigma$ either flips to another state $\sigma^\prime$ or hops to any of the four nearest neighboring sites without any restrictions. The alignment probability for particles on site $i$ is determined by the local Hamiltonian defined by Eq.~\eqref{Hapm}. If a particle on site $i$ with state $\sigma$ flips its state to $\sigma^\prime$, then the energy difference due to this flip would be:
\begin{equation}
\label{delH}
\Delta {\cal H}_i=\frac{qJ}{\rho_i}\left(n_{\sigma,i} - n_{\sigma^\prime,i} - 1\right) + q h_i\left(\delta_{\sigma,\alpha_i} - \delta_{\sigma^\prime,\alpha_i}\right) \, .
\end{equation}
Therefore, the particle makes a transition to the new state with the rate:
\begin{equation}
\label{flipeq}
    W_{\rm flip}(\sigma\to\sigma^\prime)\propto \exp\left(-\beta\Delta {\cal H}_i\right) \, ,
\end{equation}
where $\beta = T^{-1}$ is the inverse temperature. In this paper, $h_i$ is taken as a constant of sites, i.e., $h_i=h$, and $J=1$ without any loss of generality.

The hopping rate of a particle with state $\sigma$ in a direction $p$ is not affected by the external field and follows the APM hopping rule~\cite{chatterjee2020flocking,mangeat2020flocking}:
\begin{equation}\label{Whop}
    W_{\rm hop}(\sigma,p)=D\left[1+\epsilon\frac{q\delta_{\sigma,p}-1}{q-1}\right] \, .
\end{equation}
The hopping rate is $W_{\rm hop,\parallel}=D(1+\epsilon)$ for $p=\sigma$, and $W_{\rm hop,\ne}=D[1-\epsilon/(q-1)]$, otherwise. The total hopping rate for a particle is $qD$, which is independent of~$\epsilon$, then $D$ controls the particle diffusion. The mean drift velocity along the $\sigma$-direction is
\begin{equation}
v = W_{\rm hop,\parallel}-W_{\rm hop,\ne} = \frac{qD\epsilon}{q-1}.
\end{equation}
Thus, $\epsilon \in [0,q-1]$ is a dimensionless hopping-asymmetry parameter controlling the self-propulsion strength. Since $v$ is proportional to $\epsilon$ at fixed $D$ and $q$, we will often refer to $\epsilon$ as the self-propulsion parameter.

The active nature of the model originates from the coupling between the internal Potts state and particle motion. For $\epsilon>0$, particles preferentially hop along the direction associated with their internal state, resulting in a self-propulsion velocity $v$. For $\epsilon=0$, the system is no longer active ($v=0$); however, the model remains intrinsically out of equilibrium~\cite{solon2015flocking}. The reason is that although the spin-flip dynamics satisfies detailed balance with respect to the local interaction Hamiltonian ${\cal H}_i$, particle hopping is not associated with any corresponding energy change. Therefore, the combined flip-hop dynamics does not obey detailed balance and remains out of equilibrium.

Simulation evolves in the unit of Monte Carlo steps (MCS) $\Delta t$ resulting from a microscopic time $\Delta t/N$, $N$ being the total number of particles. During $\Delta t/N$, a randomly chosen particle either updates its spin state with probability $p_{\rm flip}=W_{\rm flip}\Delta t$ or hops to one of the neighboring sites with probability $p_{\rm hop}=W_{\rm hop}\Delta t$. An expression for $\Delta t$ can be obtained by minimizing the probability of nothing happens $p_{\rm wait}=1-(p_{\rm hop}+p_{\rm flip})$:
\begin{equation}\label{delt}
\Delta t=\{qD+\exp[q\beta(1+h)]\}^{-1} \, .
\end{equation}

\section{Results}\label{s3}


\subsection{Homogeneous unidirectional external field}

\begin{figure}[!t]
\centering
\includegraphics[width=\columnwidth]{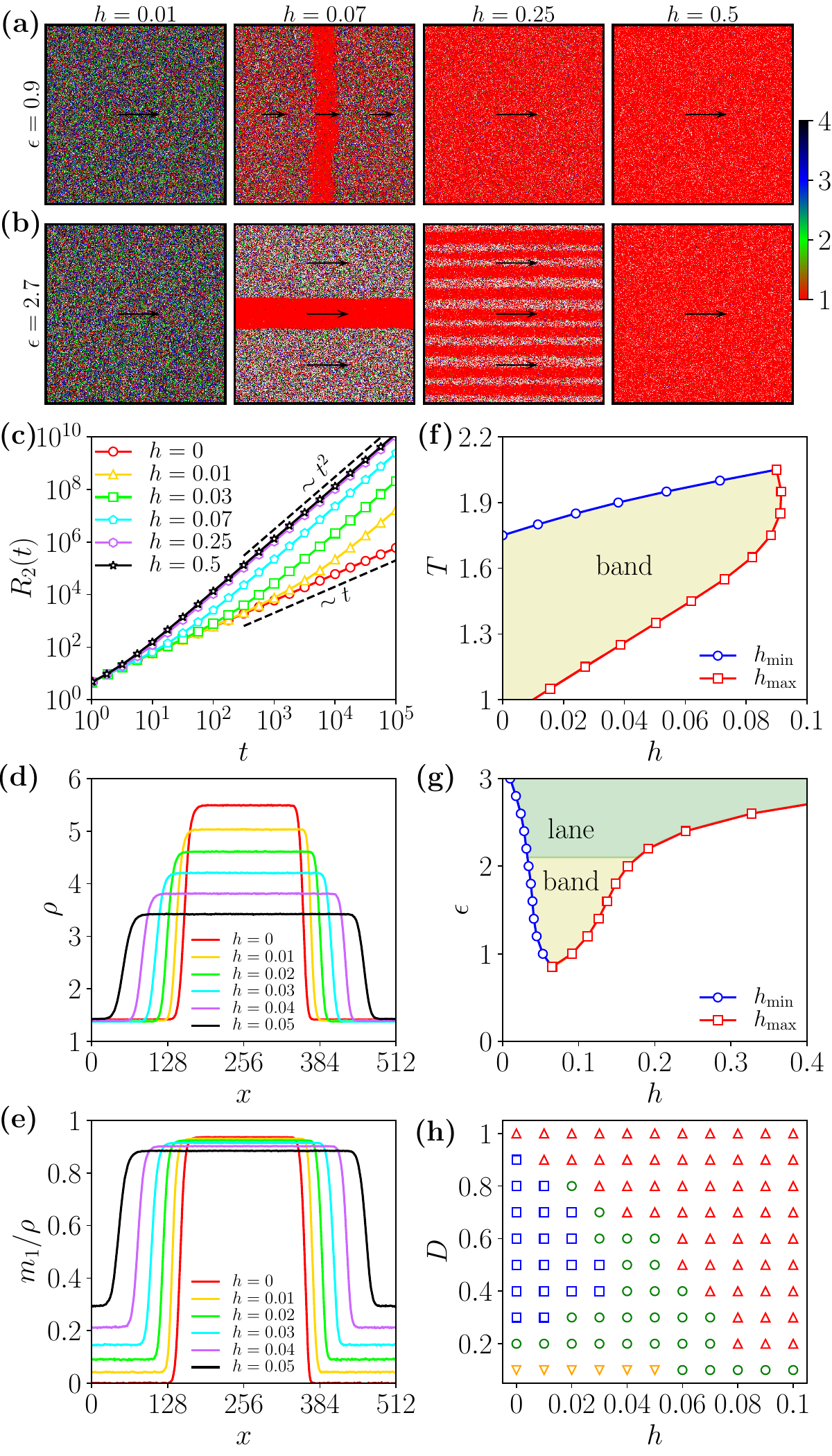}
\caption{\label{fig1}(Color online) {\it Steady-state features of 4-state APM under a homogeneous unidirectional field.} (a--b) Steady-state snapshots for increasing $h$ and (a)~$\epsilon=0.9$ and (b)~$\epsilon=2.7$, exhibiting a phase separation at intermediate $h$ values. Black arrows indicate the motion direction of the particles. Color code: red ($\sigma=1$), green ($\sigma=2$), blue ($\sigma=3$), and black ($\sigma=4$). (c)~Mean-square displacement for varying $h$ and $\epsilon=0.9$. (d--e) Time-averaged density and magnetization profiles, respectively, for $\beta=0.7$, $\epsilon=0.9$, and increasing $h$.  Parameters: $D=1$, $\beta=0.5$, $\rho_0=3$, $L=512$. (f)~$T-h$ phase diagram for $D=1$, $\epsilon=0.9$, and $\rho_0=3$. (g)~$\epsilon-h$ phase diagram for $D=1$, $\beta=0.5$, and $\rho_0=3$. (h)~$D-h$ state diagram for $\epsilon=2.7$, $\beta=1$, and $\rho_0=5$. The represented states are: polar liquid (dark-red up triangle), stripe (green circle), short-range order (blue square), and motility-induced pinning (orange down triangle).}
\end{figure}

We first analyze the behavior of the $4$-state APM under a homogeneous unidirectional external field on a square lattice of size $L=512$. The field has a constant site-independent strength ($h_i=h$) and is directed along the $\sigma=1$ state for all lattice sites $(\alpha_i=1)$, without loss of generality.

Before discussing the effects of the external field, let us briefly recall the phenomenology of the $4$-state APM in the absence of a field~\cite{chatterjee2020flocking, mangeat2020flocking}. In the liquid-gas coexistence regime, upon increasing the self-propulsion~$\epsilon$, the system undergoes a reorientation transition between two distinct phase-separated ordered states. At low~$\epsilon$, the polar liquid phase forms a {\it band}: an elongated dense liquid domain whose long axis is transverse to its direction of motion. At higher~$\epsilon$, the polar liquid phase instead forms a {\it lane}, whose long axis is parallel to its direction of motion. These two morphologies provide a reference for characterizing the field-induced states discussed below.

\subsubsection{Steady states \& phase diagrams}

We first investigate the steady state starting from either a disordered state or one aligned with the field. Figs.~\ref{fig1}(a--b) show the influence of the field $h$  on the steady state dynamics at fixed temperature ($T=2$) and density ($\rho_0=3$), for two different self-propulsion velocities: a low velocity $\epsilon=0.9$ [Fig.~\ref{fig1}(a)] and a high velocity $\epsilon=2.7$ [Fig.~\ref{fig1}(b)], starting from an initial disordered configuration. A corresponding movie (\texttt{movie1}) is available in Ref.~\cite{zenodo} showing the time evolution leading to these steady states. From Eq.~\eqref{flipeq}, flips into the field-preferred state $(\sigma'=1)$ occur at a higher rate (since $\Delta\mathcal H_i$ decreases by $q h_i$). Therefore, once the external field is applied ($h>0$), the system exhibits a local polar order along the field, analogous to equilibrium spin models under a symmetry-breaking field~\cite{chaikin1995}. Moreover, increasing the field strength $h$ further aligns the particles along the field direction, thereby enhancing the polar order (see Appendix~\ref{appA} for the non motile limit). For these parameters, the system exhibits a homogeneous polar ordered phase both at low field ($h=0.01$) with a positive global magnetization $M_1$, and at large field ($h=0.5$) where $M_1$ takes a larger value. At intermediate field values ($h=0.07$), the system exhibits a phase separation between two polar-ordered phases with different densities and magnetizations, both advecting along the field. This phase separation has the same origin as the one observed without an external field and presents the same reorientation transition~\cite{chatterjee2020flocking, mangeat2020flocking}, from a transverse band motion at low velocities [Fig.~\ref{fig1}(a)] to a longitudinal lane motion at high velocities [Fig.~\ref{fig1}(b)]. The lane motion can be constituted by one ($h=0.07$) or several ($h=0.25$) stripes, depending on the initial condition and the parameters. Indeed, increasing the external field continuously from zero in a phase-separated steady state only produces continuous changes in the densities and magnetizations of the two phases until a single polar-ordered phase is reached at a large field.

Fig.~\ref{fig1}(c) displays the mean-square displacement (MSD) of individual particles, defined as $R_2(t) = \langle |{\bf r}_k(t) - {\bf r}_k(0)|^2\rangle$, where $\mathbf{r}_k(t)$ is the position of the $k$-th particle at time $t$ and $\langle \cdot \rangle$ denotes the average over all $N$ particles, as a function of time $t$ for several field strengths. For $h=0$, the MSD scales as $R_2 \sim t$, indicating diffusive motion, characteristic of the disordered state observed for these parameters. For $h>0$, the motion becomes ballistic, with the MSD scaling as $R_2 \sim t^2$, characteristic of the polar-ordered state where the velocity increases with $h$, confirming the conclusion drawn from Figs.~\ref{fig1}(a--b). 

Figs.~\ref{fig1}(d--e) show the $y$-integrated, time-averaged density and magnetization profiles of phase-separated steady states for $\beta=0.7$, $\epsilon=0.9$, and increasing $h$. These profiles correspond to a regime already exhibiting phase separation via liquid-gas coexistence at $h=0$~\cite{chatterjee2020flocking, mangeat2020flocking}, allowing us to examine how a uniform field reshapes that coexistence. As $h$ increases, the low-density phase becomes polarized: its magnetization $m_{1}/\rho$ rises while its density stays nearly unchanged. Consequently, the standard liquid-gas flocking picture---apolar gas coexisting with a polar liquid band~\cite{solon2015flocking,chatterjee2020flocking,mangeat2020flocking}---is replaced by \textit{polar–polar phase separation}: a low-density polar background and a high-density polar band, both moving along the field. In parallel, the high-density phase slightly loses density, as the band becomes wider, and its magnetization weakens modestly. The band velocity $c$ increases linearly with $h$, from $c\simeq 1.513$ at $h=0$ to $c \simeq 1.568$ at $h=0.05$, just before the single polar-ordered phase is reached. Note that the band velocity is larger than the particle velocity $v=4D\varepsilon/3 = 1.2$.

Fig.~\ref{fig1}(f--h) presents steady-state phase diagrams showing the influence of the field strength $h$ together with, respectively, the temperature $T$, the self-propulsion velocity $\epsilon$, and the diffusion $D$. The temperature–field phase diagram [Fig.~\ref{fig1}(f)] shows that increasing the temperature shifts the phase-separation (denotes the polar–polar separation discussed before) region (in yellow) toward larger field strengths. This shift reflects the weakening of polar order at higher temperatures, which requires a stronger field to restore alignment. At $h=0$, the system exhibits a polar-ordered phase at low temperatures, a coexistence state at intermediate temperatures, and a disordered phase at high temperatures~\cite{chatterjee2020flocking, mangeat2020flocking}, which makes the phase-separation region continuous for $h>0$. At high temperatures (here $T>T_c\sim2$), no phase separation is observed, and the system always presents a homogeneous polar-ordered phase with a global magnetization that increases continuously with the field strength $h$. The critical temperature $T_c$ depends on both the density $\rho_0$ and the self-propulsion velocity $\epsilon$. This high-temperature region is analogous to the supercritical region of the liquid-gas phase transition, where the transformation from gas to liquid occurs continuously. However, its origin is more straightforward here: the regions for $h<h_{\rm min}$ and $h>h_{\rm max}$ are identical, polar-ordered phases with the same symmetries, and are therefore connected. The velocity-field phase diagram [Fig.~\ref{fig1}(g)] shows that increasing the self-propulsion velocity broadens the phase-separation region (in yellow and green) towards both lower and larger field strengths. At low velocities (here $\epsilon<\epsilon_c\sim0.9$), no phase separation is observed, corresponding to the high-temperature region discussed above. This critical velocity $\epsilon_c$ depends on both the density $\rho_0$ and the temperature $T$. At $\epsilon=\epsilon_*\sim2.1$, the reorientation transition~\cite{chatterjee2020flocking,mangeat2020flocking} occurs independently of the field strength $h$. The system exhibits a phase separation with a transverse band motion for intermediate velocities ($\epsilon_c<\epsilon<\epsilon_*$), and a longitudinal lane motion for large velocities ($\epsilon>\epsilon_*$).

Recently, it has been shown that the polar-ordered liquid phase is metastable for discrete-symmetry flocks~\cite{benvegnen2022flocking,benvegnen2023metastability,woo2024motility,mangeat2025emergent}, including the active Potts model~\cite{chatterjee2025stability}, where spontaneous droplet nucleation of transverse states can occur within the polar‐ordered phase for $D<1$, rendering it unstable and only short‐range ordered. The diffusion-field phase diagram [Fig.~\ref{fig1}(h)] shows the stability of the liquid phase under an external field, extending the set of diagrams presented in Ref.~\cite{chatterjee2025stability}. As diffusion decreases, a larger field is required for the liquid phase to remain stable. The presence of the field also affects the short-range ordered regime, promoting the formation of stripes composed of one or several lanes aligned along the field direction for $0.3 \le D \le 0.8$, close to the liquid-phase regime. Finally, for $D\le 0.1$, the phenomenon of motility-induced interface pinning appears~\cite{woo2024motility,mangeat2025emergent,chatterjee2025stability}, but is again influenced by the external field: at large $h$, the pinned interfaces evaporate, allowing the formation of stripes.

\subsubsection{Treadmilling of the longitudinal lane}

\begin{figure}[!t]
\centering
\includegraphics[width=\columnwidth]{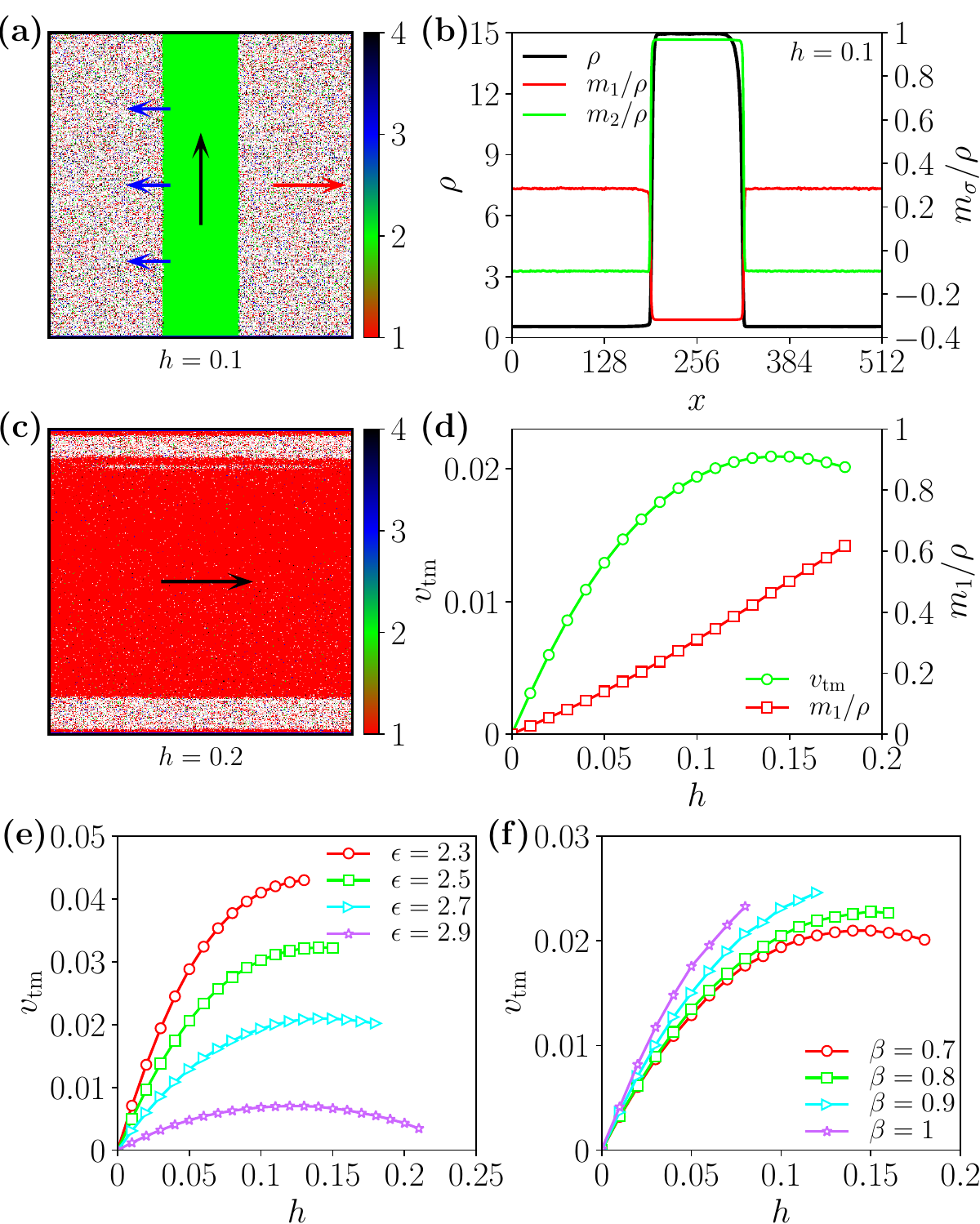}
\caption{\label{fig2}(Color online) {\it Treadmilling of the longitudinal lane under a transverse field.} (a)~Steady-state snapshot starting from a lane of state $\sigma=2$ (black arrow) for a low field strength ($h=0.1$). The lane slowly moves in the opposite direction (blue arrow) of the field (red arrow). Color code: red ($\sigma=1$), green ($\sigma=2$), blue ($\sigma=3$), and black ($\sigma=4$). (b)~Corresponding time-averaged density and magnetization profiles for state $\sigma=1$ and $\sigma=2$. The density profile is slightly tilted opposite to the field orientation. (c)~Steady-state snapshot starting from a lane of state $\sigma=2$ for a higher field strength ($h=0.2$), showing a reorientation of the lane along the field. (d)~Treadmilling velocity $v_{\rm tm}$ and maximum magnetization of the background $m_1/\rho$ as a function of the field strength $h$. $v_{\rm tm}$ reaches a plateau for large $h$ while the magnetization behaves almost linearly. Parameters: $D=1$, $\beta=0.7$, $\epsilon=2.7$, $\rho_0=4$, and $L=512$. Treadmilling velocity $v_{\rm tm}$ as a function of the field strength $h$ for: (e)~$\beta=0.7$, $\rho_0=4$, and varying $\epsilon$; and (f)~$\epsilon=2.7$, $\rho_0=\rho_*(\beta)$, and varying $\beta$. }
\end{figure}

We now investigate the steady state when the initial condition is prepared in a state ($\sigma \in \{2,3,4\}$) different from the field direction ($\alpha_i=1$), which could then be transverse to the field ($\sigma \in \{2,4\}$) or opposite to the field ($\sigma=3$). For each $\sigma$, we study two morphologies: a transverse band at small $\epsilon$ and a longitudinal lane at large $\epsilon$. The application of a weak field on a transverse band does not affect the global dynamics, while a strong field simply reorients the particles into the state $\sigma=1$, preserving the transverse motion. The same conclusion holds for the application of an opposite field on a longitudinal lane. However, a distinctive behavior emerges when a weak transverse field ($\alpha_i=1$) is applied to a longitudinal lane of state $\sigma=2$. Fig.~\ref{fig2}(a) shows the steady state for such an arrangement ($h = 0.1$), revealing a slow but steady drift (blue arrows) of the lane in the direction opposite to the field (red arrow). A corresponding movie (\texttt{movie2a}) is available in Ref.~\cite{zenodo} showing the time evolution leading to this steady state in a rectangular domain. This phenomenon can be understood from the corresponding time-averaged density and magnetization profiles shown in Fig.~\ref{fig2}(b). The weak field cannot destroy the strong order inside the longitudinal lane ($m_2>0$), but it weakly polarizes the dilute background along the field direction ($m_1>0$). Consequently, this produces an asymmetric background flux along the two lane edges: particles slowly move along the field direction, leaving the right side of the lane and accumulating on its left side, which explains the tilted density profile (black curve). This renewal-and-loss dynamics can be mechanistically interpreted as \textit{treadmilling}, where the lane advances by renewal at its left side (front) and loss at its right side (back) under the background flux. From a nonequilibrium-transport viewpoint, treadmilling is a flux-balance process sustained by addition at the front and removal at the rear. A related qualitative mechanism is well known from treadmilling in cytoskeletal filaments, where intrinsically polar actin or microtubule filaments can exhibit similar growth and shrinkage dynamics at the two ends under non-equilibrium chemical driving by ATP/GTP hydrolysis~\cite{wegner1976head, pollard2003cellular, rodionov1997microtubule, margolis1998microtubule, bugyi2010control}, and similarly in FtsZ/MreB filament dynamics in bacteria~\cite{bisson2017treadmilling,kim2006single,whitley2021ftsz}. Here, however, the analogy is only phenomenological: unlike cytoskeletal filaments, whose polarity is intrinsic and molecular, our lane has no intrinsic polarity; instead, the external field generates unequal particle fluxes through addition at one edge and removal at the opposite edge. Moreover, increasing the field strength can reorient the lane parallel to the field direction [Fig.~\ref{fig2}(c), for $h=0.2$]. The minimal field strength $h_c$ required to reorient the lane depends on the lane properties, such as its density and width. To avoid introducing this additional dependence and facilitate the analysis of the treadmilling dynamics, we initialize the system in the steady state obtained at $h=0$ and then apply the field. This choice leaves the observations of Figs.~\ref{fig2}(a–c) unaffected. We stress that the system was initialized as a lane oriented perpendicular to the external field, only to study the treadmilling mechanism in a controlled way. For a very weak field, starting from a disordered or fully ordered configuration, the system can self-organize into one or several longitudinal lanes, which subsequently undergo the same treadmilling dynamics. A movie (\texttt{movie2b}) is available in Ref.~\cite{zenodo} showing the time evolution leading to treadmilling starting from different initial conditions. Therefore, treadmilling is a robust emergent outcome of the underlying dynamics once the system self-organizes into the appropriate lane structure.

Fig.~\ref{fig2}(d) shows the treadmilling (lane-drift) velocity $v_{\rm tm}$ as a function of the field strength $h$, for $h<h_c$ ($h_c \simeq 0.19$ here). For weak fields, $v_{\rm tm}$ grows approximately linearly with $h$. In this regime, the dilute background is weakly polarized along the field, and the mean velocity scales with the polarization, $M_1 = m_1/\rho$, and the renewal flux feeding the lane edges therefore increases with $M_1$, implying that it increases with $h$. As $h$ increases further, the background keeps accelerating, but the lane also densifies, so more {\it mass} must be transported to shift it (see Appendix~\ref{appB}). This balance can be captured by the scaling law
\begin{equation}
\label{eq:vtm}
v_{\rm tm} \simeq M_1 {\cal F}( \rho_{\rm lane}/\rho_{\rm bg}, \beta)    
\end{equation}
where $\rho_{\rm bg}$ is the background density (mean density measured outside the lane), $\rho_{\rm lane}$ is the lane density, and ${\cal F}$ a function increasing with $\beta$ and decreasing with the ratio $\rho_{\rm lane}/\rho_{\rm bg}$ (see Appendix~\ref{appB}). The competing trends, $M_1$ and $\rho_{\rm lane}/\rho_{\rm bg}$ both increasing with $h$, produce a plateau in $v_{\rm tm}$ (at fixed $\beta$ and $\epsilon$) around $h \sim 0.12$ and a mild decrease as $h \to h_c$. Note that the value of $v_{\rm tm}$ is not affected by the chosen initial condition; only the value of $h_c$ depends on it. However, as in standard phase separation, the coexisting phase properties are set by the control parameters ($h$, $\beta$, $\epsilon$), and are essentially independent of the average density $\rho_0$: $\rho_0$ decides only the phase volume fraction. Consistently, both the treadmilling velocity $v_{\rm tm}$ and the reorientation threshold $h_c$ are independent of $\rho_0$, as suggested by Eq.~\eqref{eq:vtm}.

Figs.~\ref{fig2}(e--f) display the treadmilling velocity as a function of $h$ for varying $\epsilon$ and $\beta$, respectively, while keeping the initial condition in a lane configuration (large $\epsilon$ and $\beta$). In Fig.~\ref{fig2}(f), we use the density $\rho_0 = \rho_*(\beta)$ to enforce the lane configuration without affecting the results, where $\rho_*(\beta)$ is the density at which the disorder-order transition occurs for $\epsilon=0$ and $h=0$~\cite{chatterjee2020flocking, mangeat2020flocking}. Increasing $\epsilon$ (at fixed $h$) strongly reduces the treadmilling velocity and raises the reorientation threshold, as the initial lane becomes denser, reflected by the increase of the lane mass $\rho_{\rm lane}/\rho_{\rm bg}$. Conversely, increasing $\beta$ (at fixed $h$) signifies stronger polarization, enhancing the background magnetization $M_1$, while decreasing the lane mass $\rho_{\rm lane}/\rho_{\rm bg}$. As a result, $v_{\rm tm}$ increases and $h_c$ decreases. Eq.~\eqref{eq:vtm} suggests a stronger increase of $v_{\rm tm}$ with $\beta$ than can be accounted by the increase of $M_1$ and the decrease of $\rho_{\rm lane}/\rho_{\rm bg}$, due to the flipping dynamics at the back and front of the lane, which becomes faster as $\beta$ increases.


\subsection{Spatially varying orientation of the external field}

\begin{figure}[!t]
\centering
\includegraphics[width=\columnwidth]{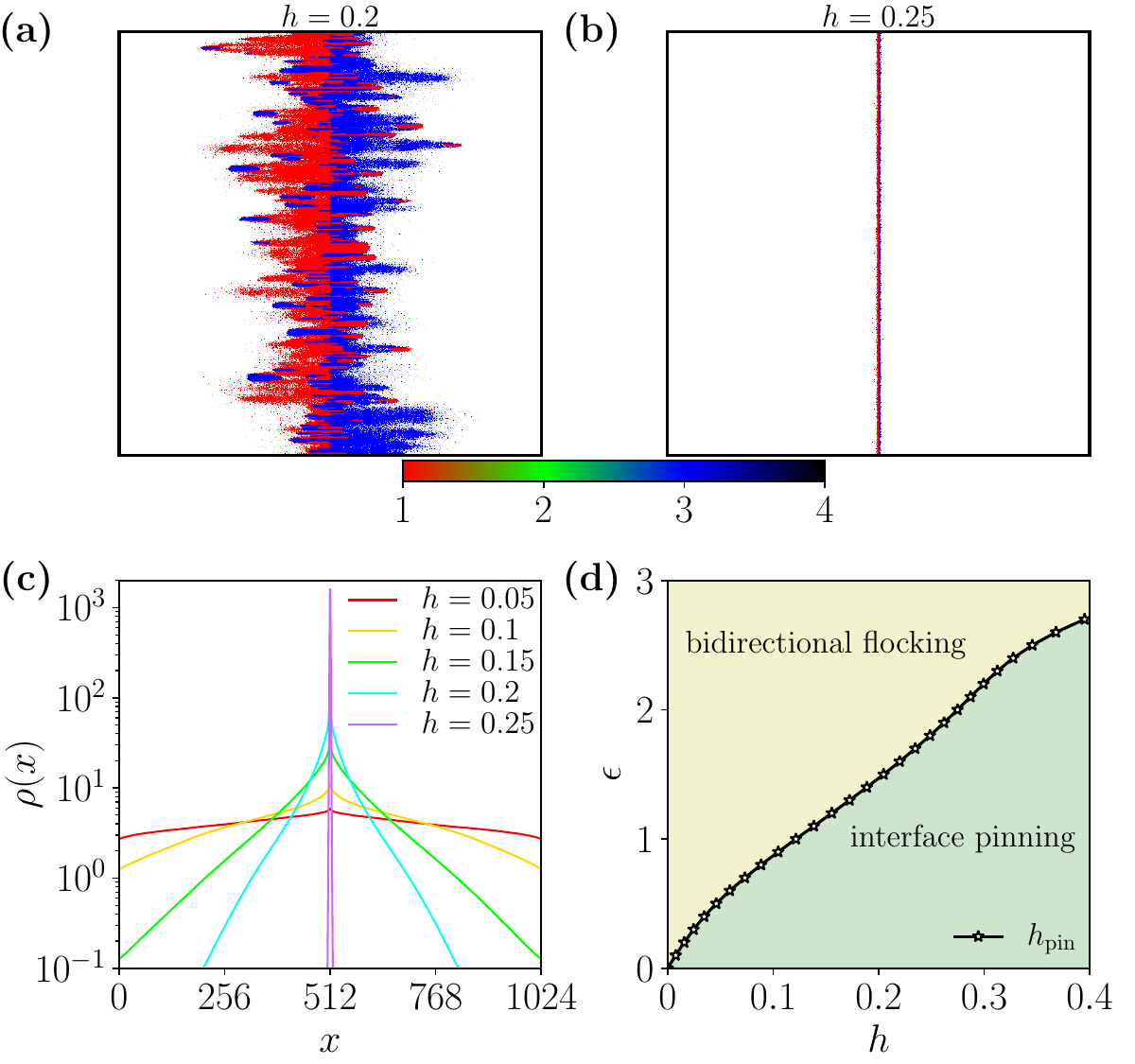}
\caption{\label{fig3}(Color online) {\it Steady-state features of 4-state APM under a bidirectional field.} (a-b)~Steady state snapshot for $\epsilon=1.8$ and increasing $h$, exhibiting the bidirectional flocking and the field-induced interface pinning (FIIP) states, respectively. Color code: red ($\sigma=1$), green ($\sigma=2$), blue ($\sigma=3$), and black ($\sigma=4$). (c)~Corresponding time-averaged density profiles for $\epsilon=1.8$ and varying $h$. (d)~$\epsilon-h$ state diagram. Parameters: $D=1$, $\beta=1$, $\rho_0=4$, $L=1024$.}
\end{figure}

We now investigate the behavior of the $4$-state APM under an external field of homogeneous magnitude ($h_i=h$) whose orientation $\alpha_i$ varies spatially across the lattice. Specifically, we consider two cases: a bidirectional orientation and a random orientation of the external field.

\subsubsection{Bidirectional orientation of external field and field-induced pinning}

We first analyze the behavior of the $4$-state APM subjected to an external field oriented in two opposite directions: rightward ($\alpha_i=1$) in the left part of the lattice ($x_i<L/2$), and leftward ($\alpha_i=3$) in the right part ($x_i \ge L/2$). Figs.~\ref{fig3}(a--b) show steady-state snapshots for $\epsilon=1.8$ and increasing $h$, starting from a disordered configuration. A corresponding movie (\texttt{movie3}) is available in Ref.~\cite{zenodo} showing the time evolution leading to these steady states. In each half of the lattice, the particles preferentially align with the field orientation: $\sigma=1$ on the left and $\sigma=3$ on the right. For a weak field, the flocking motion across the opposite-field domain is faster than the field-induced reorientation. Consequently, particles crossing the central interface do not fully realign with the local field, and the steady state exhibits coexisting right- and left-moving flocking clusters. As $h$ increases, the field-induced reorientation becomes faster, and the distance a flock travels before reorienting in the opposite-field region decreases. Therefore, flocks repeatedly switch between $\sigma = 1$ and $\sigma = 3$ around the central interface, leading to an interfacial oscillatory dynamics [Fig.~\ref{fig3}(a)]. Finally, for $h > h_{\rm pin} \simeq 0.25$, the particle orientations flip almost instantaneously upon crossing the central interface, leading to a field-induced interface pinning (FIIP) state. Particles cannot penetrate deeply into either half; instead, they accumulate at the central interface and execute a back-and-forth motion between the two interfacial columns $x_i=L/2-1$ and $x_i=L/2$ [Fig.~\ref{fig3}(b)], analogous to the motility-induced interface pinning previously reported in Refs.~\cite{woo2024motility,mangeat2025emergent,chatterjee2025stability}.

Fig.~\ref{fig3}(c) presents the $y$-integrated, time-averaged density profiles $\rho(x)$ for increasing $h$. For any nonzero field, particles retain a finite probability to flip in the opposite-field region, which effectively biases their motion back toward the central interface; consequently, all profiles develop a maximum at the interface. With increasing $h$, this interfacial peak sharpens, and the half-width $w$ of the profile around the interface---interpreted as the {\it reorientation length}, the typical distance traveled into the opposite-field region before reorienting---decreases. For $w \simeq 1$, the FIIP state is reached. For $w>1$, the system exhibits a {\it bidirectional flocking} with coexisting $\sigma=1$ and $\sigma=3$ clusters: for $1<w\lesssim L/2$ (e.g., $h=0.2$), flocks repeatedly reorient and oscillate around the central interface, whereas for $w \gtrsim L/2$ (e.g., $h=0.05$), reorientation is effectively suppressed and flocks maintain their orientation across the system. This last regime is likely restricted to $h=0$ in the thermodynamic limit, since any finite field would induce reorientation on sufficiently large length scales. Thus, for any fixed nonzero field, the regime $w \gtrsim L/2$ is a finite-size regime and cannot survive the limit $L \to \infty$. In the $L \to \infty$ limit, the dynamics remain confined to an interfacial region of width ${\cal O}(w)$ before crossing over to FIIP. The profiles in Fig.~\ref{fig3}(c) allow us to compute the velocity-field state diagram shown in Fig.~\ref{fig3}(d). As $\epsilon$ increases (at fixed $h$), the reorientation length $w$ also increases, implying that a stronger field is required to reach the FIIP state. This explains why $h_{\rm pin}$ increases with $\epsilon$. Moreover, $h_{\rm pin}$ decreases with $\beta$, as the reorientation length $w$ decreases, whereas it remains almost independent of $\rho_0$.

\begin{figure}[!t]
\centering
\includegraphics[width=\columnwidth]{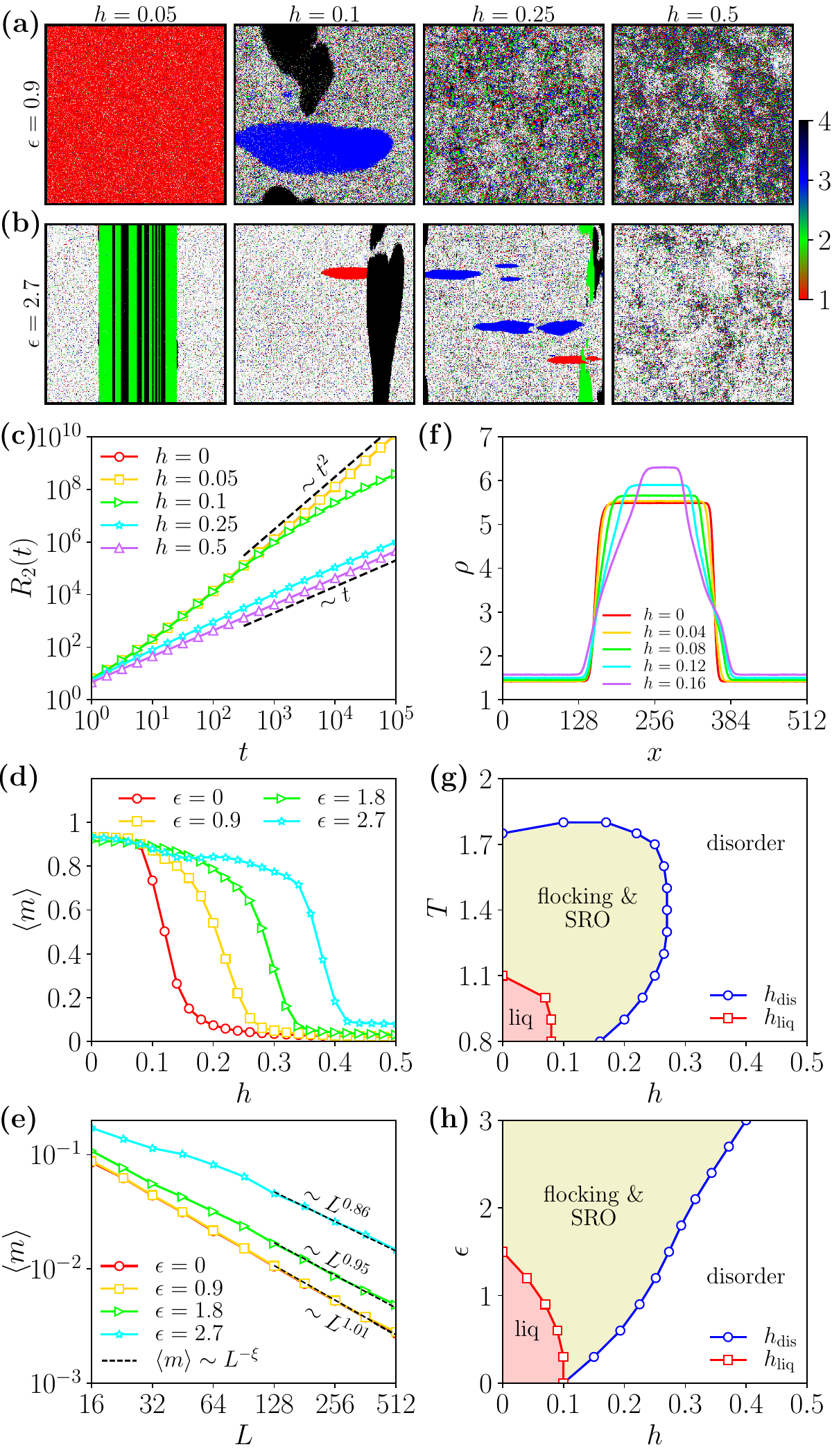}
\caption{\label{fig4}(Color online) {\it Steady-state features of 4-state APM under a random orientational field.} (a--b) Steady-state snapshots for increasing $h$ and (a)~$\epsilon=0.9$ and (b)~$\epsilon=2.7$, starting from an ordered state. Color code: red ($\sigma=1$), green ($\sigma=2$), blue ($\sigma=3$), and black ($\sigma=4$). (c)~Mean-square displacement for varying $h$ and $\epsilon=0.9$. (d)~Global magnetization $\langle m \rangle$ as a function of $h$ for several $\epsilon$ and $L=64$. (e)~Global magnetization $\langle m \rangle$ as a function of $L$ for several $\epsilon$ and $h=0.5$. Parameters: $D=1$, $\beta=1$, $\rho_0=3$, $L=512$. (f)~Time-averaged density profiles for $\beta=0.7$, $\epsilon=0.9$, and increasing $h$. (g)~$T-h$ phase diagram for $D=1$, $\epsilon=0.9$, and $\rho_0=3$. (h)~$\epsilon-h$ phase diagram for $D=1$, $\beta=1$, and $\rho_0=3$.}
\end{figure}
 
\subsubsection{Random orientation of external field}

We next study the behavior of the $4$-state APM in an external field of uniform magnitude $h_i=h$, with a site-dependent orientation $\alpha_i$ drawn independently and randomly from the four Potts directions. Figs.~\ref{fig4}(a--b) show the influence of the field $h$ on the steady state dynamics at fixed temperature ($T=1$) and density ($\rho_0=3$), for two different self-propulsion velocities: a low velocity $\epsilon=0.9$ [Fig.~\ref{fig4}(a)] and a high velocity $\epsilon=2.7$ [Fig.~\ref{fig4}(b)], starting from an initial ordered configuration oriented in state $\sigma=1$. A corresponding movie (\texttt{movie4}) is available in Ref.~\cite{zenodo} showing the time evolution leading to these steady states. Increasing the field strength $h$ locally favors an alignment of the particles along the random field orientation at each site, thereby destroying the initial liquid state, leading to a phase-separated coexistence state at intermediate field values ($h = 0.1$) and a disordered state at large field ($h = 0.5$). This disordered state consists of field-oriented sites exhibiting local order but global disorder. The particles rapidly flip into the field orientation as they traverse the lattice, leading to a FIIP in some regions (analogous to the bidirectional field), whose characteristic size decreases with $h$. The coexistence state is composed of flocking domains and FIIP-like regions. At low field, the flocking domains remain stable, with band or lane structures, whereas at higher field, the flocking domains may destabilize and reorient into a different direction due to the interaction with the FIIP-like regions. With larger self-propulsion velocities, a stronger field is required to destroy the flocking domains, thereby resulting in the disordered state, analogous to the effect of the bidirectional field. Indeed, strong self-propulsion stabilizes particle motion, reducing the pinning effect of the field. For $\epsilon=0.9$, the liquid state is stable at low fields, destabilized into a coexistence state for $0.08<h<0.23$ and into a disordered state for $h>0.23$ [Fig.~\ref{fig4}(a)]. For $\epsilon=2.7$, a lane state is stable at low fields, driven into a coexistence state for $0.06<h<0.38$ and into a disordered state for $h>0.38$ [Fig.~\ref{fig4}(b)]. 

Fig.~\ref{fig4}(c) shows the mean-squared displacement (MSD) of individual particles, $R_2(t)$, as a function of time $t$ for several field strengths, confirming the conclusion drawn from Figs.~\ref{fig4}(a--b). For $h=0$, the MSD scales as $R_2 \sim t^2$, indicating ballistic motion, characteristic of the ordered state observed for these parameters. The same ballistic scaling persists at $h=0.05$, indicating that the liquid remains stable. For $h=0.1$, the motion is ballistic at short times and becomes diffusive (with the MSD scaling as $R_2 \sim t$) at long times, corresponding to the coexistence state. For $h>0.23$, the motion becomes diffusive, typical of the disordered state for which the particles perform a random-walk-like dynamics on the lattice. Moreover, the MSD decreases with increasing $h$ in the diffusive regime, implying a decrease of the effective diffusion constant, consistent with stronger trapping/pinning and a reduced typical size of FIIP-like regions.

Fig.~\ref{fig4}(d) shows the average global magnetization $\langle m \rangle$ as a function of $h$ for several velocities, where $m$ is defined in Eq.~\eqref{eqmag} and $\langle m\rangle$ is obtained by time-averaging after relaxation and then averaging over $200$ independent realizations of the random field orientation. We observe an apparent continuous phase transition between an ordered state (liquid or coexistence state with $\langle m \rangle > 0$ at low $h$) and a disordered state ($\langle m \rangle = 0$ at large $h$) where the critical field value $h_{\rm dis}$, at which the transition occurs, monotonically increases with the self-propulsion velocity $\epsilon$. This increase of $h_{\rm dis}$ with $\epsilon$ reflects a competition between collective alignment and quenched random-field disorder: faster self-propulsion enhances the persistence of polarized domains and reduces the effective impact of spatially uncorrelated field directions, so a larger $h$ is required to destroy global order.

We emphasize that the field-driven destruction of the global order discussed in Fig.~\ref{fig4}(d) is not the same transition as the zero-field ($h=0$) order--disorder transition of the purely diffusive APM ($\epsilon=0$), which is first-order~\cite{chatterjee2020flocking, mangeat2020flocking}. In equilibrium two-dimensional systems, quenched disorder is known to round first-order transitions through the Aizenman–Wehr mechanism~\cite{aizenman1989rounding, aizenman1990rounding, cardy1999quenched}. Although the $\epsilon=0$ diffusive APM is intrinsically out of equilibrium, the quenched random-field orientation couples directly to the symmetry-breaking degree of freedom, and we indeed observe that random-field ($h>0$) suppresses the first-order signatures of the pure ($h=0$) diffusive APM~\cite{chatterjee2020flocking, mangeat2020flocking} and restores the continuous nature of the transition in the same qualitative sense as the equilibrium Aizenman--Wehr scenario. A similar extension of the equilibrium Aizenman--Wehr disorder-rounding scenario to non-equilibrium systems with absorbing states has been reported previously~\cite{villa2014quenched}. Further numerical evidence of this disorder-rounding phenomenology in the pure diffusive APM by random-field disorder is provided in Appendix~\ref{appC}.

Fig.~\ref{fig4}(e) shows the average global magnetization $\langle m\rangle$ as a function of system size $L$ in the large-field regime for several self-propulsion velocities $\epsilon$. In all cases, $\langle m\rangle$ decreases with $L$ and is well described by an algebraic finite-size scaling $\langle m\rangle \sim L^{-\xi}$. The decay of $\langle m\rangle$ with $L$ is consistent with the Imry-Ma argument~\cite{imry1975random,imry1984random,blankschtein1984potts,aizenman1989rounding} whereby quenched random fields destroy long-range order below a lower critical dimension $d_\ell$ ($d_\ell=4$ for continuous systems and $d_\ell=2$ for discrete systems~\cite{imry1984random}) by effectively breaking the system into domains whose orientations are pinned locally. Importantly, activity modifies this scaling: increasing $\epsilon$ reduces the decay exponent (e.g., from $\xi\simeq 1$ at $\epsilon=0$ to $\xi\simeq 0.86$ at $\epsilon=2.7$), indicating that self-propulsion partially counteracts disorder by allowing more persistent and coherent alignment over larger distances, even though global order still decreases with system size over the range of $L$ explored. This is in contrast to recent hydrodynamic results for incompressible polar active fluids with continuous symmetry~\cite{chen2022incompressible}, where quenched random-field disorder need not destroy coherent motion in $d>2$. In our two-dimensional discrete-symmetry RFAPM (which is a compressible active system with density fluctuations and phase-separated regions), by contrast, the same disorder is markedly more disruptive.

Fig.~\ref{fig4}(f) shows the $y$-integrated, time-averaged density profiles of phase-separated steady states for $\beta=0.7$, $\epsilon=0.9$, and increasing $h$ while keeping the field orientation fixed. These parameters already exhibit phase separation at $h=0$~\cite{chatterjee2020flocking, mangeat2020flocking}, allowing us to quantify how random orientational disorder reshapes the coexistence regime. As $h$ increases, the dilute background phase remains disordered while its density increases slightly, and the standard liquid-gas phase-separation picture remains valid. In contrast, the dense band becomes more pronounced: peak density and polarization increase as the band gets narrower. This behavior is opposite to the homogeneous unidirectional field case in Fig.~\ref{fig1}(d--e). The band velocity $c$ slightly decreases as $h$ increases, from $c\simeq 1.514$ at $h=0$ to $c\simeq1.486$ at $h=0.2$, just before the band is fully destroyed. It remains larger than the particle velocity $v=4D\varepsilon/3 = 1.2$.

Figs.~\ref{fig4}(g--h) summarize the steady-state behavior in the presence of a random field by the $T-h$ and $\epsilon-h$ phase diagrams, respectively. In the temperature–field diagram [Fig.~\ref{fig4}(g)], the ordered liquid persists at low $T$ and weak disorder. Increasing either $T$ or $h$ drives the system into a coexistence regime where flocking domains coexist with FIIP-like regions. Depending on parameters, these flocking domains may appear as long-lived band or lane structures~\cite{chatterjee2020flocking, mangeat2020flocking} or as transient short-range–ordered (SRO) flocks~\cite{chatterjee2025stability}. The stability of band and lane structures are system-size dependent, leading to SRO flocks in the thermodynamic limit. For larger $T$ and/or $h$, the system becomes fully disordered. The transition to the disordered state, denoted by $h_{\rm dis}$ in Fig.~\ref{fig4}(g), is determined from the Binder cumulant of the global magnetization $m$, $U_4 = 1-3\langle m^4\rangle/5\langle m^2\rangle^2$, where the prefactor corresponds to an effective three-component (tetrahedral) Potts order parameter and ensures $U_4\to0$ in a Gaussian disordered phase; $h_{\rm dis}$ is then identified by the crossing of $U_4(h)$ for different system sizes $L$. At high temperatures (here $T>T_c\simeq 1.8$), global polar order is absent ($\langle m\rangle\simeq 0$) throughout, and increasing $h$ produces only a smooth crossover from the homogeneous gas at $h=0$ to a random-field--disordered state for $h>0$. This critical temperature $T_c$ depends on both the density $\rho_0$ and the self-propulsion velocity $\epsilon$. The velocity–field diagram [Fig.~\ref{fig4}(h)] shows an analogous sequence: at low $\epsilon$ and weak disorder, the ordered liquid is stable; increasing $\epsilon$ or $h$ produces a coexistence regime (flocking domains plus FIIP-like regions), followed by a fully disordered state at sufficiently large $h$. Consistent with Fig.~\ref{fig4}(d), the disordering threshold $h_{\rm dis}$ shifts to larger values as $\epsilon$ increases.


\subsection{Hydrodynamic description}

\begin{figure*}[!t]
\centering
\includegraphics[width=\textwidth]{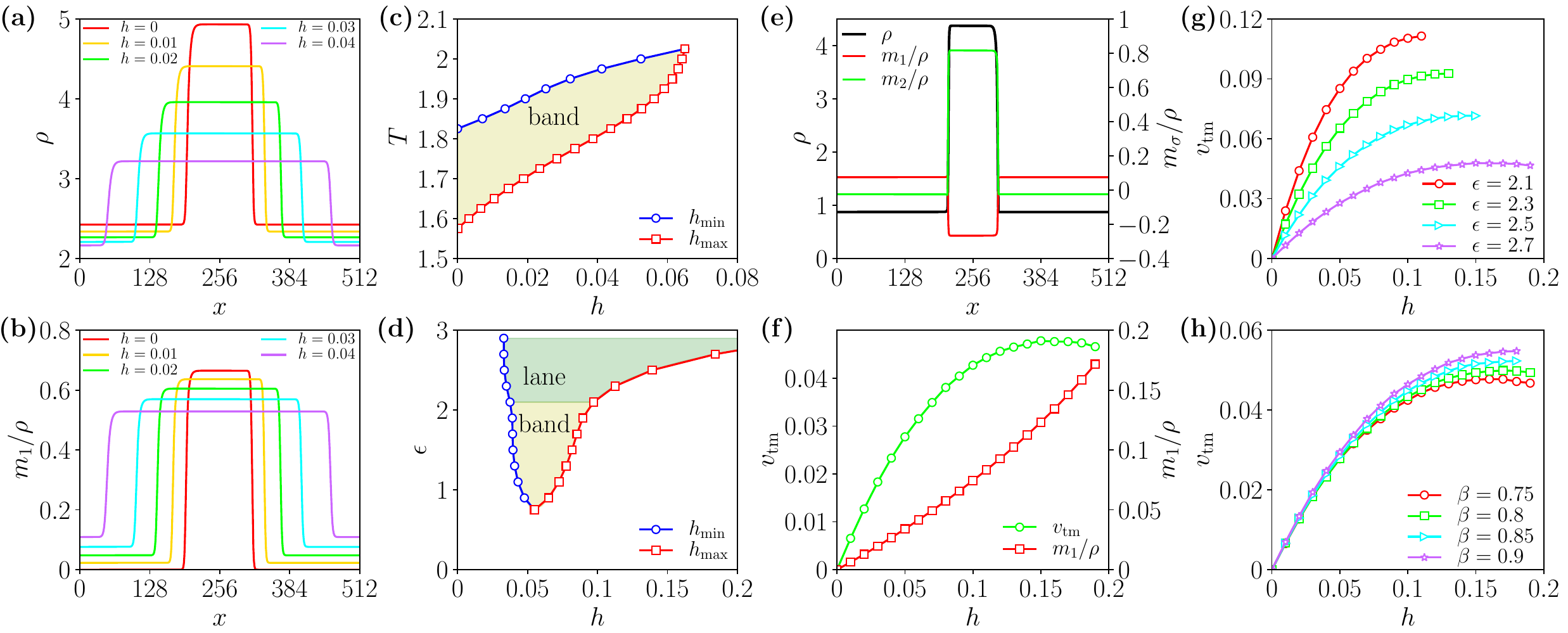}
\caption{\label{fig5}(Color online) {\it Hydrodynamic theory under a homogeneous unidirectional field.} (a--b)~Steady state density and magnetization profiles for $\beta=0.55$, $\epsilon=0.9$, $\rho_0=3$, $L=512$, and increasing $h$. (c)~$T-h$ phase diagram for $\rho_0=3$ and $\epsilon=0.9$. (d)~$\epsilon-h$ phase diagram for $\rho_0=3$ and $\beta=0.5$. (e)~Steady-state density and magnetization profiles for state $\sigma=1$ and $\sigma=2$. The density profile is slightly tilted opposite to the field orientation. (f)~Treadmilling velocity $v_{\rm tm}$ and maximum magnetization of the background $m_1/\rho$ as a function of the field strength $h$. $v_{\rm tm}$ reaches a plateau for large $h$ while the magnetization behaves almost linearly. Parameters: $\beta=0.75$, $\epsilon=2.7$, $\rho_0=1.5$, and $L=512$. (g)~Treadmilling velocity $v_{\rm tm}$ for varying $\epsilon$ for $\beta=0.75$ and $\rho_0=1.5$. (h)~Treadmilling velocity $v_{\rm tm}$ for varying $\beta$ for $\epsilon=2.7$ and $\rho_0=\rho_*(\beta)$.}
\end{figure*}

We now present our results of the coarse-grained hydrodynamic description of the APM with an external field, defined as $h_\sigma = h \delta_{\sigma,\alpha}$, which is equal to $h$ in the direction $\sigma = \alpha$, and $0$ in the other directions. If we define the average density of particles in the state $\sigma$ at the two-dimensional position {\bf x} as $\rho_\sigma({\bf x},t) = \langle n_{\sigma,i}(t) \rangle$, the hydrodynamic equation of the density field $\rho_\sigma$ for state $\sigma$ is given by:
\begin{equation}
\label{PDEhydro0}
\partial_t \rho_\sigma = D_\parallel \partial_\parallel^2 \rho_\sigma + D_\perp \partial_\perp^2 \rho_\sigma - v \partial_\parallel \rho_\sigma + e^{r/2\rho} \sum_{\sigma' \ne \sigma } {\cal I}_{\sigma \sigma'},
\end{equation}
where the interaction term ${\cal I}_{\sigma \sigma'}$ is
\begin{align}
{\cal I}_{\sigma \sigma'} &= \left(\rho_{\sigma} + \rho_{\sigma'} - \frac{r}{4\beta J}\right) \sinh \left[ \frac{4\beta J}{\rho}(\rho_{\sigma}-\rho_{\sigma'}) + 4 \beta \Delta h\right]  \nonumber \\
&- (\rho_{\sigma}-\rho_{\sigma'}) \cosh \left[ \frac{4\beta J}{\rho}(\rho_{\sigma}-\rho_{\sigma'}) + 4 \beta \Delta h\right],
\end{align}
as derived in Appendix~\ref{appD} and Refs.~\cite{chatterjee2020flocking, mangeat2020flocking}. $D_{\parallel}=D(1+\epsilon/3)$ and $D_{\perp}=D(1-\epsilon/3)$ are the diffusion constants in the parallel ${\bf e_\parallel}$ and perpendicular ${\bf e_\perp}$ directions. The self-propulsion velocity $v=4D\epsilon/3$ is along ${\bf e_\parallel}$. Derivatives in these directions are denoted as $\partial_\parallel = {\bf e_\parallel} \cdot \nabla$ and $\partial_\perp = {\bf e_\perp} \cdot \nabla$. The field difference $\Delta h = h_\sigma - h_{\sigma'} = h (\delta_{\sigma,\alpha}-\delta_{\sigma',\alpha})$ takes value in $0$ and $\pm h$.

The hydrodynamic equations~\eqref{PDEhydro0} are integrated numerically using the Euler forward time-centered space (FTCS) differencing scheme~\cite{press2007numerical}. We solve the system of four coupled partial differential equations, one for each internal state $\sigma$, defined on a two-dimensional square domain of size $L \times L$, with periodic boundary conditions imposed along both spatial directions. Simulations are performed with $L=512$ up to a final time $t_{\rm sim}=10^4$. Numerical stability is ensured by choosing spatial and temporal discretizations $\Delta x=0.25$ and $\Delta t=10^{-3}$, which satisfy the Courant-Friedrichs-Lewy stability condition~\cite{courant1928partiellen}. We fix the parameters $D=J=r=1$, thereby setting the units of time, temperature, and density. The system is initialized with a high-density band or lane centered in the domain on a low-density gaseous background.

\begin{figure}[!t]
\centering
\includegraphics[width=\columnwidth]{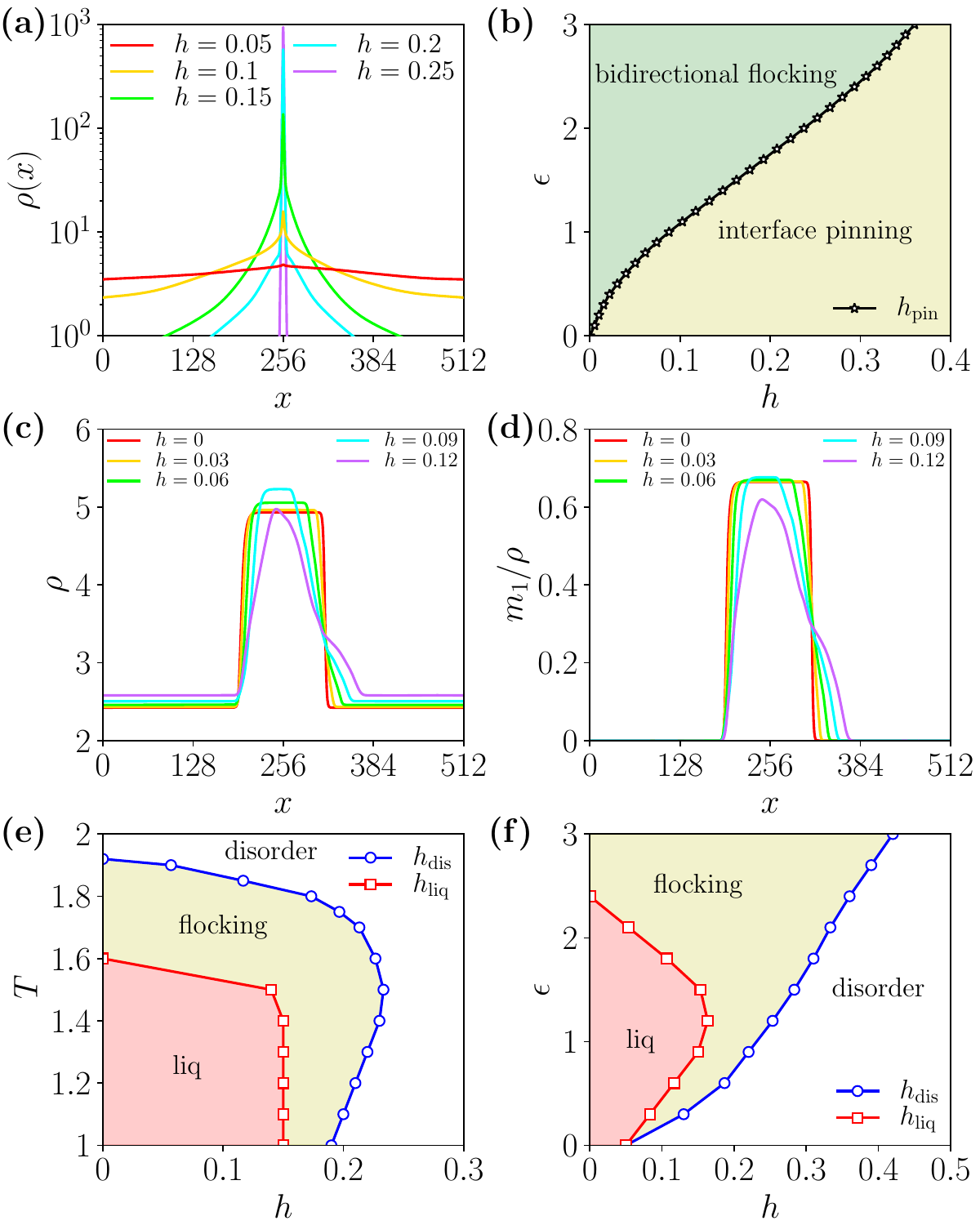}
\caption{\label{fig6}(Color online) {\it Hydrodynamic theory under (a--b)~bidirectional and (c--f)~random orientational fields.} (a)~Time-averaged density profiles for $\epsilon=1.8$ and varying $h$. (b)~$\epsilon-h$ state diagram. Parameters: $\beta=1$, $\rho_0=4$, $L=512$. (c--d)~Time-averaged density and magnetization profiles for $\beta=0.55$, $\epsilon=0.9$, $\rho_0=3$, $L=512$, and increasing $h$. (e)~$T-h$ phase diagram for $\rho_0=3$ and $\epsilon=0.9$. (f)~$\epsilon-h$ phase diagram for $\rho_0=3$ and $\beta=0.8$.}
\end{figure}

Fig.~\ref{fig5} and Fig.~\ref{fig6} present numerical solutions of the coarse-grained hydrodynamic equations and shows that the continuum theory reproduces and validates the principal steady-state structures and field-controlled interfacial phenomena observed in the microscopic simulations. Fig.~\ref{fig5}(a--b) shows the steady-state density and magnetization profiles under a homogeneous unidirectional field [cf. Fig.~\ref{fig1}(d--e)]. As the field strength $h$ increases, the profiles evolve in the same manner as in the particle model: the coexistence state remains phase separated, but the dense region and the dilute background are both field-aligned, with a stronger polarization in the dense phase and a weaker polarization in the background. This confirms, within hydrodynamics theory, the field-driven restructuring of coexistence from the zero-field liquid--gas picture toward coexistence between two polar phases with different densities and polarizations.

The corresponding $T-h$ and $\epsilon-h$ phase diagrams [Fig.~\ref{fig5}(c--d)] recover the same phase-sequence and state-diagram topology seen in the microscopic simulations [cf. Fig.~\ref{fig1}(f--g)]: a finite field interval bounded by $h_{\rm min}$ and $h_{\rm max}$ within which phase-separated flocking states persist, together with the band/lane reorientation and its systematic dependence on temperature and self-propulsion. In this sense, the hydrodynamic theory captures the same field-controlled coexistence structure and orientational reorganization mechanisms as the microscopic model.

Fig.~\ref{fig5}(e--h) provides the hydrodynamic realization of the treadmilling lane under a transverse field [cf. Fig.~\ref{fig2}]. The density and magnetization profiles in Fig.~\ref{fig5}(e) exhibit the same interfacial polarization mismatch identified in the particle simulations: a strongly ordered dense lane coexisting with a weakly polarized background, with a small asymmetry (tilt) in the density profile opposite to the imposed field direction [cf. Fig.~\ref{fig2}(b)]. This asymmetry is the continuum signature of the treadmilling drift. Fig.~\ref{fig5}(f) further shows that the treadmilling velocity $v_{\rm tm}$ increases at small $h$ and then saturates to a broad plateau at larger $h$, while the background magnetization grows approximately linearly [cf. Fig.~\ref{fig2}(d)]. The same nontrivial separation between transport and polarization trends [see Eq.~\eqref{eq:vtm}] is therefore reproduced by the coarse-grained theory (see Appendix~\ref{appB}). The parametric dependences in Fig.~\ref{fig5}(g-h), obtained by varying $\epsilon$ and $\beta$, also follow the same qualitative picture and trend as in the microscopic simulations [cf. Fig.~\ref{fig2}(e--f)], demonstrating that the hydrodynamic equations capture not only the existence of treadmilling but also its systematic control by activity and alignment strength.

Fig.~\ref{fig6}(a--b) reproduces the bidirectional-field interfacial phenomenology [cf. Fig.~\ref{fig3}(c--d)]. The time-averaged density profiles in Fig.~\ref{fig6}(a) develop an increasingly sharp accumulation peak at the field-reversal boundary as $h$ increases, indicating a reduction of penetration depth and a crossover from bidirectional flocking to a pinned interfacial state. A movie (\texttt{movie5}) is available in Ref.~\cite{zenodo} showing the time evolution of two-dimensional solutions. The $\epsilon-h$ diagram in Fig.~\ref{fig6}(b) summarizes this crossover through a pinning threshold $h_{\rm pin}$, which shifts to larger values with increasing self-propulsion, again in agreement with the microscopic simulations.

Having validated the hydrodynamic description for homogeneous unidirectional and bidirectional fields, we next examine the effect of a quenched random orientational disorder in the external field. In the continuum description, the orientation of the external field is defined as a piecewise-constant function of space to faithfully represent the quenched field disorder of the microscopic model. Specifically, the field orientation is taken to be constant as $\alpha(x,y) = \alpha_i$ within unit cells defined as $\lfloor x \rfloor \le x < \lfloor x \rfloor + 1$ and $\lfloor y \rfloor \le y < \lfloor y \rfloor + 1$, where $\alpha_i$ denotes the orientation assigned in the microscopic model and $\lfloor x \rfloor$ represents the integer part of $x$. This construction ensures that the spatially varying field remains independent of the numerical discretization and therefore admits a well-defined continuum limit as $\Delta x \rightarrow 0$.

Using this representation, the hydrodynamic equations capture the influence of quenched orientational disorder on the density fields. The steady-state profiles shown in Fig.~\ref{fig6}(c--d) demonstrate that increasing the field strength progressively disrupts the coherent flocking structures and reduces the global polarization, consistent with the behavior observed in the microscopic simulations with random-field orientations [cf. Fig.~\ref{fig4}(f)]. A corresponding movie (\texttt{movie6}) is available in Ref.~\cite{zenodo} showing the time evolution leading to these steady-state profiles. For weak fields, the system remains in a flocking polar band state, whereas increasing the field strength induces a crossover toward a disordered regime characterized by locally polarized but globally uncorrelated regions. The corresponding $T$--$h$ and $\epsilon$--$h$ phase diagrams, shown in Fig.~\ref{fig6}(e--f), summarize this transition. In agreement with the particle simulations [cf. Fig.~\ref{fig4}(g-h)], the hydrodynamic theory predicts three regimes: an ordered flocking liquid for weak fields, an intermediate coexistence regime where flocking domains coexist with short-range ordered structures, and a fully disordered state at sufficiently strong random fields. The disordering threshold $h_{\mathrm{dis}}$ increases with self-propulsion, reflecting the competition between collective alignment and quenched orientational disorder.

\section{Summary and Outlook} \label{s4}

External fields are among the most effective means of rendering active matter programmable. However, in many flocking models, they primarily act as a symmetry-breaking bias that smoothly aligns an already existing ordered phase~\cite{marchetti2013hydrodynamics, bechinger2016active, gompper2025}. Here, by coupling a minimal active Potts flock~\cite{chatterjee2020flocking,mangeat2020flocking} to a weak field, we observe a qualitatively stronger effect: even a small bias can reorganize phase coexistence and, importantly, the interfacial transport that sustains macroscopic patterns. Across homogeneous, bidirectional, and random field protocols, a common feature emerges: the field does not simply select a direction of motion; it alters how particles are converted, renewed, and trapped at interfaces, giving rise to new non-equilibrium steady states.

In the motile regime under a homogeneous, weak, unidirectional field---where the field bias remains perturbative and does not dominate the orientation dynamics---the system self-organizes, when in the phase-separated regime, into a high-density polar lane whose long axis is perpendicular to the field. Remarkably, this lane \textit{treadmills against the field direction}, with a steady drift velocity within the two-phase coexistence regime, although its magnitude varies systematically with field strength. The underlying cause is a polarization mismatch across the interface: the weak field polarizes the dilute background and drives a slow flux along the field, while the dense lane remains polarization-locked by strong local order. This mismatch produces asymmetric gain/loss at the two lane sides (renewal at the front, erosion at the back), translating the lane opposite to the field. The mechanism is qualitatively reminiscent of treadmilling in cytoskeletal filaments~\cite{wegner1976head, pollard2003cellular, bisson2017treadmilling}, where intrinsically polar filaments undergo treadmilling, although in our case the effect is purely phenomenological and arises due to an external field. To our knowledge, this treadmilling phenomenon, where transport is dominated by conversion at interfaces rather than bulk advection, has no counterpart in standard dry-flocking models and illustrates how even a weak directional bias can generate qualitatively new spatiotemporal organization.

In a bidirectional-field protocol, where the field points in opposite directions in the two halves of the system, the interface becomes an active boundary layer where field-driven reorientation competes with flock advection. At weak fields, clusters can penetrate deep into the opposite-field region before turning around, leading to a steady state with coexisting right- and left-moving flocks ({\it bidirectional flocking}). As the field is increased, reorientation accelerates, and the penetration depth decreases, until the dynamics crosses over to a \textit{field-induced interface pinning} regime in which particles accumulate at the interface, and we identify a distinct interfacial {\it oscillatory flocking state}. This arrested-interface state is field-induced and is distinct from motility-induced pinning mechanisms reported previously~\cite{woo2024motility,mangeat2025emergent,chatterjee2025stability}. This is significant because it shows that the design of the spatial field can switch an active system from sustained bulk transport to an interfacially trapped steady state.

Introducing quenched random-field orientations leads to a spatially mixed steady state, where flocking domains coexist with localized interface-pinned regions. The random field plays a dual role: it locally biases polarization and thereby nucleates pinned regions, while at the same time acting as quenched disorder that destabilizes global polar order as the system size increases, in the spirit of the Imry--Ma (IM) argument~\cite{imry1984random,blankschtein1984potts}.  In the diffusive limit, increasing quenched field disorder suppresses the sharp first-order signatures of the zero-field transition and replaces them with a smoother, continuous-like transition, in the same qualitative spirit as the equilibrium Aizenman--Wehr disorder-rounding scenario~\cite{aizenman1989rounding,aizenman1990rounding,villa2014quenched}. Importantly, this occurs in an active non-equilibrium system: activity counteracts random-field-induced disorder without invalidating these disorder-driven organizing mechanisms (IM and AW). Increasing self-propulsion shifts the disordering threshold to higher field strengths and reduces the decay of global order with system size, i.e., activity renormalizes both the loss of global order and the disorder-rounded crossover. Furthermore, the strong agreement between the microscopic simulations and the coarse-grained hydrodynamic theory indicates that the observed field-induced effects are not finite-size artifacts and supports their robustness in the thermodynamic (long-wavelength) limit.

Our results point to several promising directions for future work, centered on two complementary field-controlled interfacial mechanisms: \textit{interfacial transport} (treadmilling) and \textit{interfacial trapping} (FIIP). On the experimental side, the treadmilling state may be realized in systems such as light-activated active colloids~\cite{palacci2013living} and field-driven colloidal rollers or microrollers with tunable propulsion under electric or rotating magnetic fields~\cite{bricard2013emergence,driscoll2017unstable}. The FIIP state provides a clear example of a traveling flock transforming into a localized, oscillatory, high-density "active barrier". In active transport settings, this mechanism could be used to trap or redirect particles using only a patterned field, i.e., a switchable gate that does not require fabricating walls or obstacles. More generally, these two mechanisms suggest a concrete control strategy: by tuning the spatial structure of the field, one can switch the system between a translating macroscopic state (treadmilling) and a pinned interfacial state (FIIP), enabling programmable transport and arrest. Another natural extension is to study two-dimensional continuous-symmetry, compressible Vicsek-type flocks in a quenched random-field landscape, where spatial variations in the field may compete with polar alignment to disrupt, reshape, or pin propagating structures. This direction is motivated by earlier works showing that such systems can remain robust against quenched disorder, exhibiting quasi-long-ranged order in two dimensions~\cite{chepizhko2013optimal,toner2018swarming,das2018polar} and coherent motion persisting even under random-field disorder in incompressible active fluids~\cite{chen2022incompressible,chen2022packed}. Finally, we expect these results to motivate further studies aimed at controlling self-organization in both synthetic active systems and biological collectives.

\section*{Acknowledgements}

RP and MK thank the Indian Association for the Cultivation of Science (IACS) for the computational facility during the initial phase of this work. SC acknowledges support from the German Research Foundation (DFG) via SFB 1027 during the phase of this work carried out at Saarland University. MM and HR are financially supported by the German Research Foundation (DFG) within the Collaborative Research Center SFB 1027. 

\section*{Author contributions}

MK, SC, and RP designed the research. MK and MM performed the numerical simulations. MK, MM, and SC analysed the results and wrote the paper. HR and RP provided supervision and critical feedback. All authors reviewed and approved the final manuscript.

\section*{Data availability}

The data that support the findings of this article are publicly available~\cite{zenodo2}.

\clearpage


\appendix

\begin{figure}[tbp]
\centering
\includegraphics[width=\columnwidth]{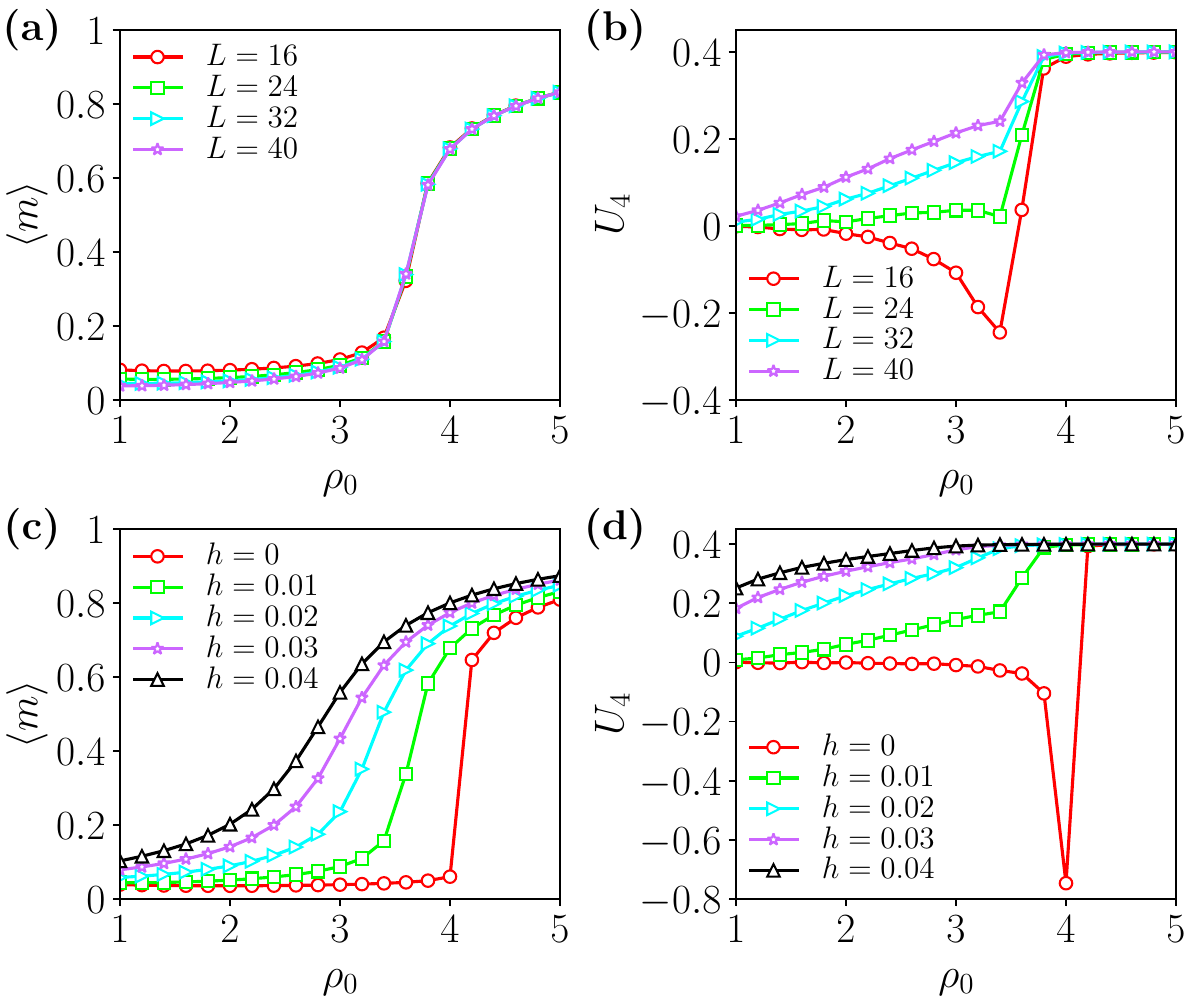}
\caption{(Color online) {\it Non-motile APM under a homogeneous unidirectional field.} (a--b) Global magnetization $\langle m \rangle$ and Binder cumulant $U_4$ for fixed $h=0.01$ and varying $\rho_0$ and $L$. There is no crossing for different $L$ in the $U_4$, suggesting no phase transition. (c--d) $\langle m \rangle$ and $U_4$ for fixed $L=32$ and varying $\rho_0$ and $h$. Parameters: $D=1$, $\beta=0.6$, and $\epsilon=0$. \label{fig7}}
\end{figure}

\section{Non-motile APM under a homogeneous unidirectional field}
\label{appA}

Fig.~\ref{fig7} characterizes the non-motile APM ($\epsilon=0$) under a homogeneous unidirectional external field, providing the diffusive reference case for the field-induced ordering discussed in Fig.~\ref{fig1}. Figs.~\ref{fig7}(a--b) show the average global magnetization $\langle m\rangle$, with $m$ defined by Eq.~\eqref{eqmag}, and the Binder cumulant $U_4=1-3\langle m^4 \rangle / 5 \langle m^2 \rangle^2$ as functions of the mean density $\rho_0$ for fixed field strength $h=0.01$ and several system sizes $L$. Although $\langle m\rangle$ increases with density, the $U_4$ curves do not exhibit a crossing, indicating the absence of a genuine phase transition. Instead, the uniform field acts as an explicit symmetry-breaking field that continuously biases the particles toward the preferred direction.

Figs.~\ref{fig7}(c--d) show $\langle m\rangle$ and $U_4$ for fixed system size $L=32$ and increasing field strength $h$. For $h=0$, the curves recover the standard first-order-like signatures of the pure diffusive APM: a sharp increase of $\langle m\rangle$ across the transition region together with a pronounced negative dip in $U_4$~\cite{chatterjee2020flocking,mangeat2020flocking}. For any finite homogeneous field $h>0$, however, these signatures are smoothed out: the magnetization rises continuously and $U_4$ changes without developing the finite-size features of a sharp transition. Thus, in the non-motile limit, a homogeneous unidirectional field does not generate a separate ordering transition, but rather continuously enhances the polar order by explicitly favoring one Potts direction. This is fully consistent with the behavior described in Fig.~\ref{fig1}, where the homogeneous field locally aligns particles along its direction and stronger fields further enhance the resulting polar order.

\begin{figure}[tbp]
\centering
\includegraphics[width=\columnwidth]{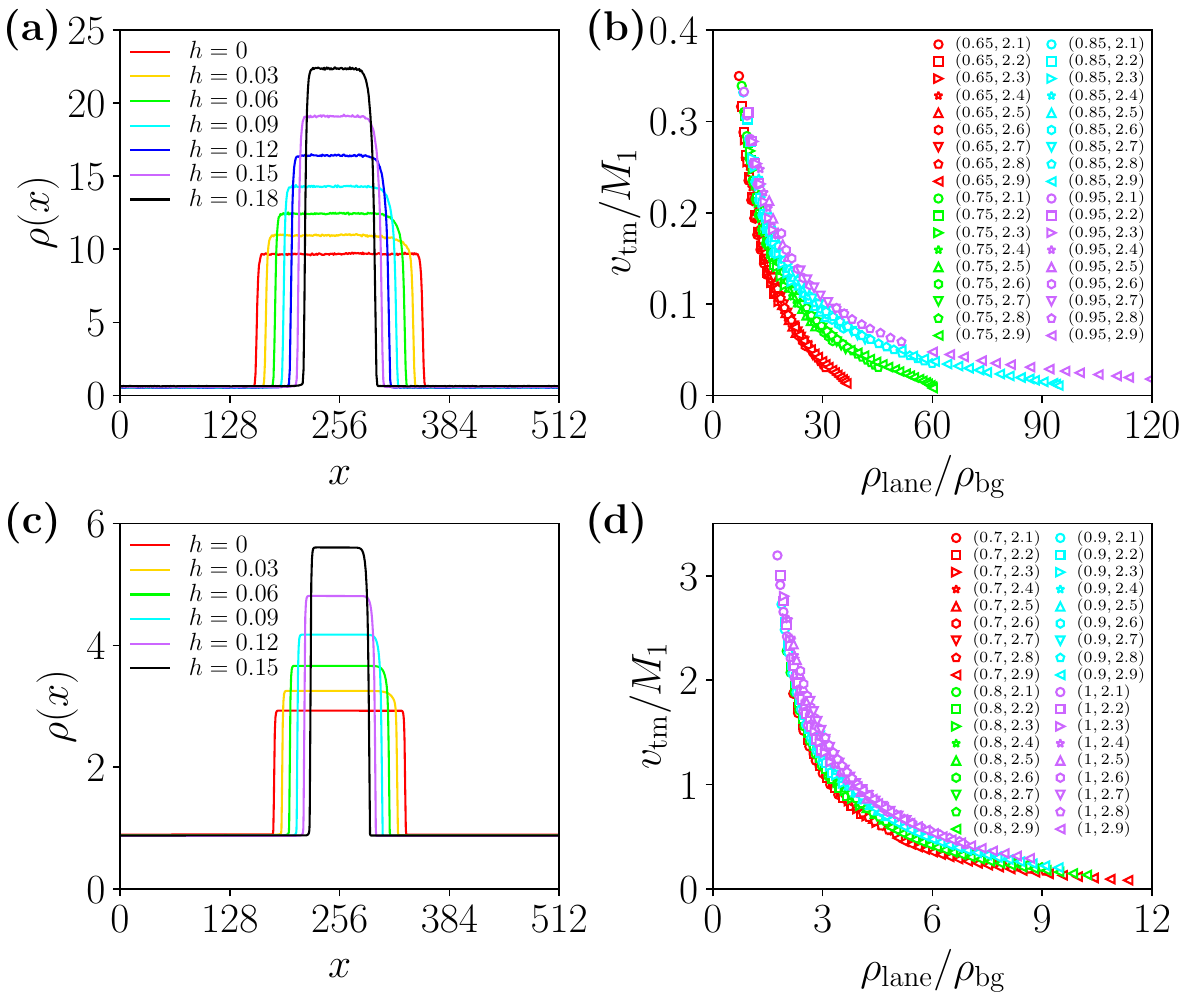}
\caption{(Color online) {\it Lane profiles and treadmilling velocity under a transverse field}. Results for (a--b)~numerical simulations and (c--d)~hydrodynamic theory. (a) \& (c)~Density profile of the lane for $\beta=0.7$, $\epsilon=2.7$ and increasing $h$. (b) \& (d)~Treadmilling velocity for several $(\beta,\epsilon)$ parameters as a function of $\rho_{\rm lane}/\rho_{\rm bg}$. \label{fig8}}
\end{figure}

\section{Lane profiles and treadmilling velocity under a transverse field}
\label{appB}

Fig.~\ref{fig8} provides the mechanistic support for the treadmilling behavior reported in Fig.~\ref{fig2}. Fig.~\ref{fig8}(a) and Fig.~\ref{fig8}(c) show the time-averaged lane-density profiles under a transverse field for $\beta=0.7$ and $\epsilon=2.7$, obtained respectively from particle simulations and hydrodynamic theory. As the field strength $h$ increases within the treadmilling regime, the lane becomes progressively narrower and denser, so that the ratio $\rho_{\rm lane}/\rho_{\rm bg}$ increases. Thus, while the field enhances the polarization of the dilute background and therefore the flux feeding the lane edges, it also increases the lane mass that must be transported to shift the lane.

Fig.~\ref{fig8}(b) and Fig.~\ref{fig8}(d) quantify this competition by plotting the normalized treadmilling velocity $v_{\rm tm}/M_1$ as a function of $\rho_{\rm lane}/\rho_{\rm bg}$ for several $(\beta,\epsilon)$ values, where $M_1=m_1/\rho$ denotes the field-induced background polarization. The data show that $v_{\rm tm}/M_1$ decreases systematically as $\rho_{\rm lane}/\rho_{\rm bg}$ increases, demonstrating that denser lanes are harder to treadmill. At fixed lane-mass ratio, the curves retain a residual dependence on $\beta$, consistent with faster interfacial flipping dynamics at larger $\beta$. Fig.~\ref{fig8}, therefore, substantiates the scaling form of Eq.~\eqref{eq:vtm} with ${\cal F}$ decreasing with $\rho_{\rm lane}/\rho_{\rm bg}$ and increasing with $\beta$. The qualitative agreement between Fig.~\ref{fig8}(a,b) and Fig.~\ref{fig8}(c,d) further shows that the hydrodynamic theory captures not only the existence of treadmilling but also the transport balance underlying the plateau of $v_{\rm tm}$ observed in Fig.~\ref{fig2}(d\text{--}f).

\begin{figure}[tbp]
\centering
\includegraphics[width=\columnwidth]{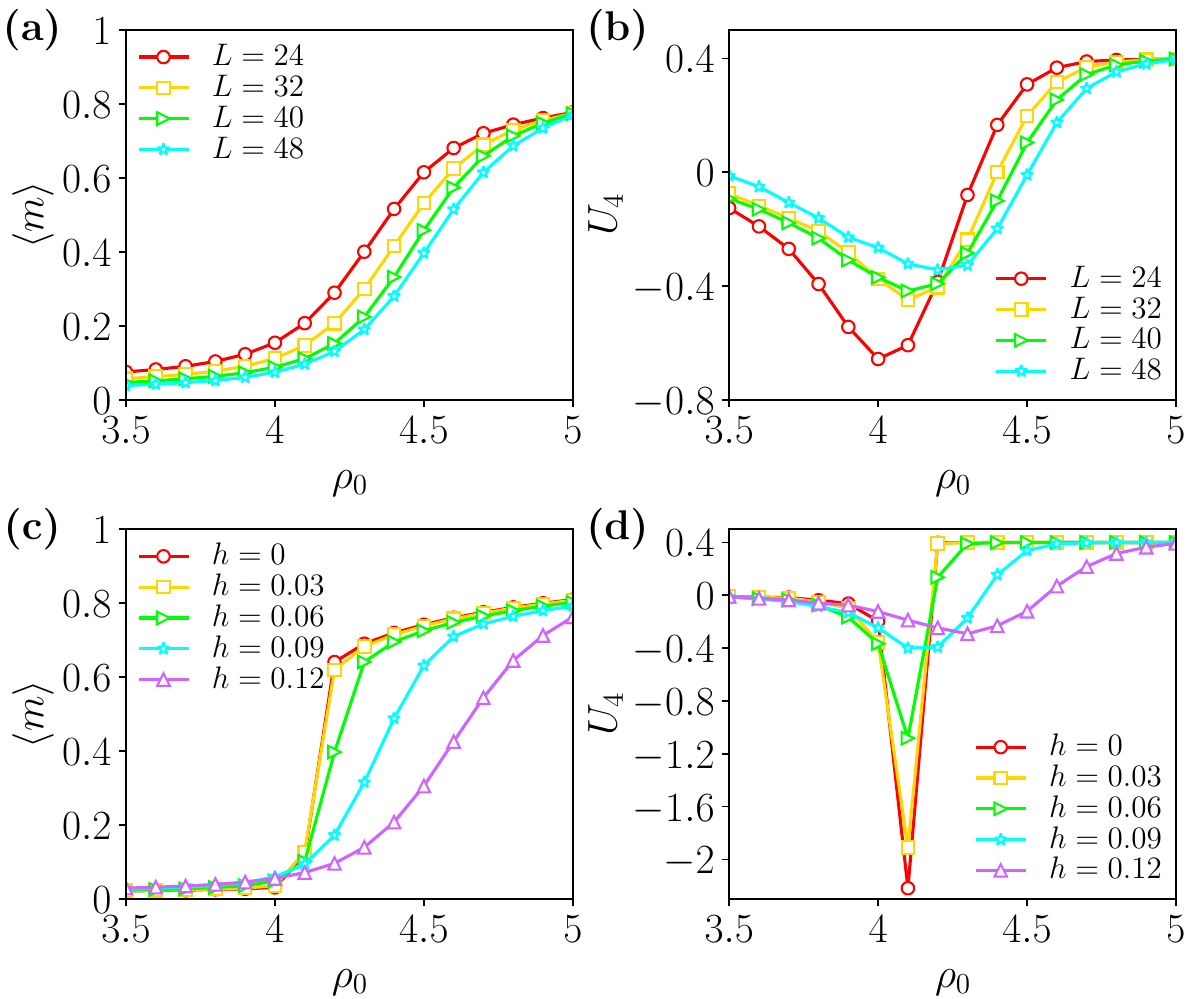}
\caption{(Color online) {\it Non-motile APM under a random orientational field.} (a--b) Global magnetization $\langle m \rangle$ and Binder cumulant $U_4$ for fixed $h=0.12$ and varying $\rho_0$ and $L$. The transition takes place at $\rho_0 \sim 4.25$. (c--d) $\langle m \rangle$ and $U_4$ versus $\rho_0$ for fixed $L=64$ and varying $h$. Parameters: $D=1$, $\beta=0.6$, and $\epsilon=0$. \label{fig9}}
\end{figure}

\section{Non-motile APM under a random orientational field}
\label{appC}

Fig.~\ref{fig9} provides the numerical evidence that in the non-motile limit $\epsilon=0$, a quenched random orientational field rounds the first-order order-disorder transition~\cite{aizenman1989rounding,aizenman1990rounding,villa2014quenched} of the pure diffusive APM~\cite{chatterjee2020flocking,mangeat2020flocking}. Figs.~\ref{fig9}(a--b) show the average global magnetization $\langle m\rangle$, with $m$ defined by Eq.~\eqref{eqmag}, and the corresponding Binder cumulant $U_4 = 1-3\langle m^4\rangle/5\langle m^2\rangle^2$, as functions of the mean density $\rho_0$ for a fixed random-field strength $h=0.12$ and several system sizes $L$. The transition occurs around $\rho_0 \sim 4.25$. As $L$ increases, $\langle m\rangle$ rises smoothly with density, while $U_4(L)$ does not develop a sharp common crossing. Instead, the negative dip in $U_4$ becomes progressively weaker with increasing system size, rather than deeper as expected for phase coexistence at a first-order transition. This finite-size trend shows that, at nonzero random-field strength, the system does not recover the discontinuous behavior of the pure diffusive APM, but instead, the quenched random field suppresses the first-order signatures and rounds the transition into a continuous-like one.

This conclusion is further supported by Figs.~\ref{fig9}(c--d), which show $\langle m\rangle$ and $U_4$ for fixed system size $L=64$ and increasing random-field strength $h$. In the absence of disorder ($h=0$), $\langle m\rangle$ exhibits a sharp increase across the transition region and $U_4$ develops a pronounced negative dip, which are the standard finite-size signatures of the first-order transition of the pure diffusive model~\cite{chatterjee2020flocking,mangeat2020flocking}. As $h$ is increased, the rise of $\langle m\rangle$ becomes progressively smoother, and the negative minimum of $U_4$ is strongly reduced. At the same time, the apparent ordering threshold shifts to larger densities, showing that a higher mean density is required for collective alignment to overcome the local pinning induced by the quenched random field. Thus, although Fig.~\ref{fig4}(d) concerns the field-driven destruction of order in the active system, Fig.~\ref{fig9} demonstrates convincingly that in the diffusive limit, random-field disorder converts the underlying transition from first-order to continuous-like, fully consistent with the Aizenman--Wehr scenario in two dimensions~\cite{aizenman1989rounding, aizenman1990rounding}.

\section{Derivation of hydrodynamic equations}
\label{appD}

Consider the $q$-state active Potts model (APM) with an external field, defined as $h_\sigma = h \delta_{\sigma,\alpha}$, which is equal to $h$ in the direction $\sigma = \alpha$, and $0$ in the other directions. From the flipping and hopping rules, we can derive the master equation defining the dynamic equation for the number of particles $n_{\sigma,i}(t)$ on site $i$ in state $\sigma$:
\begin{align}
\langle n_{\sigma,i}(t+dt) \rangle &= \left\langle n_{\sigma,i}(t) \left[ 1 - dt \sum_p W_{\rm hop}(\sigma,p) \right.\right.\nonumber\\
& \qquad\qquad\quad\left.\left.- dt \sum_{\sigma' \ne \sigma} W_{\rm flip}(\sigma \to\sigma') \right] \right\rangle \nonumber\\
&+ \left\langle \sum_p n_{\sigma,i-p}(t) W_{\rm hop}(\sigma,p) dt\right\rangle \nonumber\\
&+ \left\langle\sum_{\sigma' \ne \sigma } n_{\sigma',i}(t) W_{\rm flip}(\sigma'\to\sigma) dt \right\rangle,
\end{align}
where a subscript $i+p$ denotes the neighbor of site $i$ in the $p$-direction. In the limit $dt \rightarrow 0$, this expression yields
\begin{align}
\label{MasterEq0}
\partial_t \langle n_{\sigma,i} \rangle &= \sum_{p} \left\langle W_{\rm hop}(\sigma,p) \left( n_{\sigma,i-p}  -  n_{\sigma,i} \right)  \right\rangle \nonumber \\
&~+ \sum_{\sigma' \ne \sigma } \left\langle n_{\sigma',i} W_{\rm flip}(\sigma' \to \sigma) - n_{\sigma,i} W_{\rm flip}(\sigma\to\sigma') \right\rangle, \nonumber \\
&= \langle I_{\rm hop} \rangle + \sum_{\sigma' \ne \sigma } \langle I_{\sigma \sigma'} \rangle.
\end{align}

With Eq.~(\ref{Whop}) for the hopping rates $W_{\rm hop}$, the hopping term can be expressed as
\begin{align}
I_{\rm hop} &= \sum_{p} W_{\rm hop}(\sigma,p) \left(  n_{\sigma,i-p} -  n_{\sigma,i}  \right) \nonumber\\
&= D_0 \sum_p \left( n_{\sigma,i-p}  -  n_{\sigma,i} \right) + v \left( n_{\sigma,i-\sigma} - n_{\sigma,i} \right),
\end{align}
where the first term corresponds to a jump in a random direction $p$ with $D_0 = D[1-\epsilon/(q-1)]$ and the second term to a jump in the favored direction $\sigma$ with $v=qD\epsilon/(q-1)$. We take the hydrodynamic limit for small lattice spacing $1/L$, corresponding to large system size. We define the density of particles in the state $\sigma$ at the 2d position ${\bf x}$ as $\rho_\sigma({\bf x},t) = n_{\sigma,i}(t)$ for which the coordinate ${\bf x}$ matches the lattice site $i$ at integer positions. Let ${\bf e_p}$ be the unitary vector in the direction $p$. The Taylor expansion of $\rho_\sigma({\bf x} + {\bf e_p},t) $ gives
\begin{align}
n_{\sigma,i+p} &= \rho_\sigma({\bf x} + {\bf e_p},t) \nonumber \\
&= \rho_\sigma({\bf x},t) + \frac{\partial \rho_\sigma}{\partial p} ({\bf x},t) + \frac{1}{2} \frac{\partial^2 \rho_\sigma}{\partial p^2} ({\bf x},t) + \cdots
\end{align}
for the derivative $ \frac{\partial}{\partial p} = {\bf e_p} \cdot \nabla_{\bf x}$. We can then expand the two terms in the hopping term $I_{\rm hop}$ involving the particle number difference (for $q>2$) on neighboring sites:
\begin{equation}
n_{\sigma,i-\sigma} - n_{\sigma,i} = - \frac{\partial \rho_\sigma}{\partial \sigma}({\bf x},t) + \frac{1}{2} \frac{\partial^2 \rho_\sigma}{\partial \sigma^2} ({\bf x},t),\label{expConv}
\end{equation}
and
\begin{align}
\sum_p \left[ n_{\sigma,i-p} - n_{\sigma,i} \right] &= \sum_p \left[ -\frac{\partial \rho_\sigma}{\partial p} ({\bf x},t) + \frac{1}{2} \frac{\partial^2 \rho_\sigma}{\partial p^2} ({\bf x},t) \right] \nonumber\\
&= \frac{q}{4}\nabla_{\bf x}^2 \rho_\sigma({\bf x},t).
\end{align}
The hopping term then writes
\begin{equation}
I_{\rm hop} =  D_\parallel \partial_\parallel^2 \rho_\sigma + D_\perp \partial_\perp^2 \rho_\sigma - v \partial_\parallel \rho_\sigma,
\end{equation}
where $\partial_\parallel = \partial_\sigma$ is the derivation in the parallel direction ${\bf e_\parallel}={\bf e_\sigma}$ and $\partial_\perp$ is the derivative in the orthogonal direction ${\bf e_\perp}$ (${\bf e_\perp} \cdot {\bf e_\sigma} =0$). The diffusion constants are then $D_\parallel = qD_0/4 + v/2 = qD[1+\epsilon/(q-1)]/4$ and $D_\perp = qD_0/4 = qD[1-\epsilon/(q-1)]/4$ in the parallel and the perpendicular directions, respectively. The hydrodynamic equation for the average density of particles $\langle \rho_\sigma \rangle$ writes then
\begin{equation}
\partial_t \langle \rho_\sigma \rangle = D_\parallel \partial_\parallel^2 \langle \rho_\sigma \rangle + D_\perp \partial_\perp^2 \langle \rho_\sigma \rangle - v \partial_\parallel \langle \rho_\sigma \rangle + \sum_{\sigma' \ne \sigma } \langle I_{\sigma \sigma'} \rangle.
\end{equation}

In terms of the particle densities $\rho_\sigma$, the flipping rate writes
\begin{equation}
W_{\rm flip}(\sigma \to \sigma') =\exp\left( - \frac{q\beta J}{\rho}\Delta\rho - q \beta \Delta h\right),
\end{equation}
where $\Delta\rho=\rho_\sigma - \rho_{\sigma'}$ and $\Delta h = h_\sigma - h_{\sigma'}$ can take three values: $+h$ if $\alpha=\sigma$, $-h$ if $\alpha=\sigma'$, or $0$ if $\alpha \ne \sigma,\sigma'$. The flipping term is then equal to
\begin{align}
I_{\sigma \sigma'} &=\rho_{\sigma'} W_{\rm flip}(\sigma' \to \sigma) - \rho_\sigma W_{\rm flip}(\sigma\to\sigma') \nonumber \\
&= \Sigma\rho \sinh \left( \frac{q\beta J}{\rho}\Delta\rho + q \beta \Delta h\right) \nonumber \\
&\qquad - \Delta\rho \cosh \left( \frac{q\beta J}{\rho}\Delta\rho + q \beta \Delta h\right),
\end{align}
with $\Sigma\rho=\rho_\sigma + \rho_{\sigma'}$. We perform the transformation $\rho_\sigma \to m_\sigma$ such that $m_\sigma = (q\rho_\sigma - \rho)/(q-1)$, meaning $\Delta \rho = (q-1)(m_\sigma-m_{\sigma'})/q$ and $\Sigma\rho = 2\rho/q + (q-1)(m_\sigma+m_{\sigma'})/q$.
\begin{align}
I_{\sigma \sigma'} &= \left( \frac{2\rho}{q} + \frac{q-1}{q}\Sigma m\right) \sinh \left[ \frac{(q-1)\beta J}{\rho}\Delta m + q \beta \Delta h\right] \nonumber \\
&- \frac{q-1}{q}\Delta m \cosh \left[ \frac{(q-1)\beta J}{\rho}\Delta m + q \beta \Delta h\right],
\end{align}
where $\Delta m = m_\sigma-m_{\sigma'}$ and $\Sigma m = m_\sigma+m_{\sigma'}$. We consider that $m_\sigma$ are independent and identically distributed Gaussian variables with mean $\langle m_\sigma \rangle$ and variance $\sigma^2 = \alpha_m \langle \rho \rangle$. We may rewrite $I_{\sigma \sigma'}$ in terms of $X=\Sigma m$ and $Y=\Delta m$, leading to
\begin{align}
I_{XY} &= \left( \frac{2\rho}{q} + \frac{q-1}{q}X\right) \sinh \left( BY + H\right) \nonumber \\
&- \frac{q-1}{q}Y \cosh \left( BY + H\right),
\end{align}
with $B=(q-1) \beta J/\rho$, and $H=q\beta \Delta h$. We define then
\begin{equation}
\langle I_{XY} \rangle = \int dX P(X)\int dY P(Y) I_{XY},
\end{equation}
with
\begin{equation}
\quad P(X) = \frac{1}{\sqrt{4\pi \sigma^2}} \exp[-(X-X_0)^2/4\sigma^2],
\end{equation}
since the variance of $X$ and $Y$ are $2\sigma^2$, and with $X_0=\langle m_\sigma \rangle + \langle m_{\sigma'}\rangle = \langle\Sigma m\rangle $ and $Y_0=\langle m_\sigma \rangle - \langle m_{\sigma'} \rangle = \langle\Delta m\rangle$. Performing these Gaussian integrals, we get
\begin{align}
\langle I_{XY} \rangle &= \left\{ \left[ \frac{2\rho}{q} + \frac{q-1}{q}X_0 - \frac{q-1}{q}(2B\sigma^2)\right] \sinh \left( BY_0 + H\right) \right. \nonumber  \\
&\left.- \frac{q-1}{q}Y_0 \cosh \left( BY_0 + H\right) \right\} \exp(B^2 \sigma^2).
\end{align}
Replacing $B$, $H$, and $\sigma$ by their values and going back to the $\rho_\sigma$ variable, we obtain
\begin{align}
\langle I_{\sigma \sigma'} \rangle &= \left[\left(\langle\Sigma\rho\rangle - \frac{r}{q\beta J}\right) \sinh \left( \frac{q\beta J}{\langle\rho\rangle}\langle\Delta\rho\rangle + q \beta \Delta h\right) \right. \nonumber \\
&\left.- \langle\Delta\rho\rangle \cosh \left( \frac{q\beta J}{\langle\rho\rangle}\langle\Delta\rho\rangle + q \beta \Delta h\right) \right] \exp(r/2\rho),
\end{align}
where $r=2 (q-1)^2(\beta J)^2 \alpha_m$ has been chosen to fit the perturbative theory when $\langle\Delta\rho\rangle \ll \langle\rho\rangle$ and $h =0$:
\begin{align}
\langle I_{\sigma \sigma'} \rangle = \left(\frac{q\beta J}{\langle\rho\rangle} \langle\Sigma\rho\rangle -1 - \frac{r}{\langle\rho\rangle} \right) \langle\Delta\rho\rangle + {\cal O}(\langle\Delta\rho\rangle^3).
\end{align}



\bibliography{rfapm_refs}

\end{document}